\documentclass[reprint,aps,prl,floatfix,amsmath,amssymb,amsfonts]{revtex4-2}

\usepackage{bm}
\usepackage{graphicx,color}
\usepackage{xcolor}
\usepackage{physics}
\usepackage{dsfont}
\usepackage[normalem]{ulem}
\usepackage{hyperref,url}
\usepackage{mathrsfs}
\usepackage{calrsfs}
\usepackage{bbold}
\usepackage{mathtools}
\usepackage{comment}
\usepackage{scalerel}
\usepackage{tikz}

\newcommand*{\blue}{\textcolor{blue}}

\newcommand{\mathbfs}[1]{\boldsymbol{\mathbf{#1}}}

\usepackage{subfig}

\DeclareMathOperator{\diag}{diag}
\DeclareMathOperator{\ci}{ci}

\renewcommand{\vec}[1]{\mathbf{#1}}

\usepackage{varioref}

\usepackage[active]{srcltx}

\expandafter\ifx\csname package@font\endcsname\relax\else
\expandafter\expandafter
\expandafter\usepackage
\expandafter\expandafter
\expandafter{\csname package@font\endcsname}%
\fi
\hypersetup{
	colorlinks=true,       
	linkcolor=blue,          
	citecolor=blue,        
}

\usepackage[section]{placeins}
\usepackage{enumitem}
\usepackage{fancybox}
\labelformat{figure}{Fig.\,#1}

\DeclareMathAlphabet{\mathcal}{OMS}{cmsy}{m}{n}
\newcommand{\be}{\begin{equation}}
\newcommand{\ee}{\end{equation}}
\newcommand{\bi}{\begin{enumerate}}
\newcommand{\ei}{\end{enumerate}}

\begin{document}
    \title{Emergence of volume-law scaling for entanglement negativity from the 
     Hawking radiation of analogue black holes}  

\author{S. Mahesh Chandran} 
\email{maheshchandran@snu.ac.kr}
\affiliation{Seoul National University, Department of Physics and Astronomy, Center for Theoretical Physics, Seoul 08826, Korea}
\author{Uwe R. Fischer} 
\email{uwerf@snu.ac.kr}
\affiliation{Seoul National University, Department of Physics and Astronomy, Center for Theoretical Physics, Seoul 08826, Korea}
%

\begin{abstract}
The quantum information content of Hawking radiation holds the key to understanding black-hole evaporation and the fate of unitarity. Motivated by recent advances in cold-atom experiments, we develop a lattice-regularization approach aimed at simulating the coarse-grained entanglement scaling of a quantum field in a 1+1D analogue black-hole background. We provide the first concrete demonstration that logarithmic negativity --- an entanglement monotone that typically exhibits a UV-divergent log-scaling for the conformal  
vacuum --- acquires a UV-finite volume term from the nonlocal correlations seeded by Hawking radiation. We show that this volume term encodes the number density as well as the spatial distribution of entangled Hawking pairs along the black-hole interior and exterior. 
We highlight its prospective detection in currently realizable experiments 
and its implications beyond the analogue paradigm, in particular for black-hole thermodynamics. 
\end{abstract}

\maketitle

\textit{Introduction.} 
Hawking's seminal prediction that black holes radiate~\cite{1975HawkingCiMP} raises fundamental questions about how information is encoded in the emitted quanta, and whether unitarity is ultimately preserved~\cite{2017Unruh.WaldRoPiP}. 
The key lies in the entanglement content of Hawking radiation (HR), whose late-time behavior carries decisive imprints of unitarity via the Page curve~\cite{1993PagePRLa,2013PageJoCaAP,2021Almheiri.etalRMP,2020Almheiri.etalJoHEP}. 
Extracting entanglement measures in quantum field theory (QFT) is however notoriously subtle, as they are plagued by ultraviolet (UV) divergences that can obscure physically relevant correlations. 
Accordingly, most approaches adopt a regularization prescription to compute the entanglement between spatial bipartitions of quantum fields, primarily in terms of entanglement entropy~\cite{1985tHooftNPB,1986Bombelli.etalPRD,1993SrednickiPRL,1994Callan.WilczekPLB,1994Holzhey.etalNPB,2004Calabrese.CardyJoSMTaE,2004Casini.HuertaPLB,2009Casini.HuertaJoPAMaT,2018Sorkin.YazdiCaQG,2026Jones.Yazdi}. These reveal an \textit{area-law} scaling arising from short-distance vacuum correlations characteristic of gapped quantum systems~\cite{2010Eisert.etalRMP}, and entailing direct implications for black-hole entropy~\cite{1997Mukohyama.etalPRD,1998Mukohyama.etalPRD,1998Mukohyama.IsraelPRD,2008Das.etal,2008Das.etal,2011SolodukhinLRiR,2020Chandran.ShankaranarayananPRD,2025Belfiglio.etalPRD}. 
In contrast, typical pure states exhibit a \textit{volume-law} scaling of entanglement~\cite{1993PagePRL,2022Bianchi.etalPQ}, motivating the question of how correlations generated by pair-creation processes, such as HR, are spatially organized. These considerations 
have also been generalized for mixed-state entanglement measures, in particular logarithmic negativity~\cite{2002Audenaert.etalPRA,2016Eisler.ZimborasPRB,2021Shapourian.etalPQ}, which we employ in this work. Despite promising insights from various related settings~\cite{1999Jacobson.MattinglyPRD,2006Das.ShankaranarayananPRD,2007Das.ShankaranarayananCaQG,2007Jacobson.ParentaniPRD}, the entanglement scaling for 
global states that incorporate Hawking correlations remains unresolved.



In recent years, cold-atom experiments have enabled the tomographic reconstruction of quantum states via correlation measurements~\cite{2015Steffens.etalNC,2020Gluza.etalCP,2025Murtadho.etalPRR}, allowing the extraction of scaling laws for information-theoretic measures~\cite{2023Tajik.etalNP,2025Jarema.etala,2025Jarema.etal}. Concurrently, the field of analogue gravity has facilitated 
laboratory access to curved spacetime QFT~\cite{1981UnruhPRL,2001Barcelo.etalCaQG,2003Barcelo.etalPRA,2003Fedichev.FischerPRL,2004Fedichev.FischerPRA,2017Chae.FischerPRL,2018Eckel.etalPRX,2022Viermann.etalN,2016SteinhauerNP,2019MunozdeNova.etalN,2021Kolobov.etalNP,2011Barcelo.etalLRiR,2023Braunstein.etalNRP,2025SchuetzholdPiPaNP}, leading to 
the landmark observation of Hawking pair production via smoking gun correlation  signatures~\cite{2008Balbinot.etalPRA,2008Carusotto.etalNJoP,2013Anderson.etalPRD,2016SteinhauerNP,2019MunozdeNova.etalN,2021Kolobov.etalNP}. These developments suggest that the entanglement scaling of HR — long confined to theoretical investigations — may now be within direct 
experimental reach. 
Predicting the scaling structure, however, requires a regularization scheme that goes beyond 
existing techniques.

In this Letter, we introduce a lattice-regularization framework that, for the first time, enables the systematic extraction of entanglement scaling associated with HR, from an experimentally accessible analogue black-hole model. Leveraging the Painlevé-Gullstrand (PG) description of acoustic black holes in a quasi-one-dimensional (quasi-1D)  Bose-Einstein condensate (BEC), we demonstrate that the logarithmic negativity --- 
which typically yields a UV-divergent log-scaling corresponding to the conformal vacuum~\cite{2012Calabrese.etalPRL} --- acquires a UV-finite volume term for the Unruh state~\cite{1976UnruhPRD}, as a consequence of the nonlocal correlations seeded by HR. This volume law emerges outside the \textit{quantum atmosphere}~\cite{2016GiddingsPLB} (the near-horizon region dominated by vacuum correlations), with a slope that is sensitive to both the surface gravity and the propagation velocities of Hawking quanta inside and outside the horizon.

Our results establish that the negativity scaling of HR carries a distinct, regulator-independent signature of the pair-creation process that can be detected in currently realizable analogue setups. 
We provide the complete expression for this scaling (including prefactors), and demonstrate its relation to the number density of entangled Hawking pairs and their spatial distribution across the black-hole horizon. Beyond the analogue setup, our regularization scheme provides a general tool for extracting finite entanglement measures from QFT in curved spacetimes, with immediate applications to black-hole thermodynamics~\cite{1997Mukohyama.etalPRD,1998Mukohyama.etalPRD,1998Mukohyama.IsraelPRD,2008Das.etal,2008Das.etal,2011SolodukhinLRiR,2020Chandran.ShankaranarayananPRD,2025Belfiglio.etalPRD}, gravitational collapse~\cite{2022Balbinot.FabbriPRD,2025Belfiglio.etalPRDa,2025Sahota.etalPRD}, and early-universe cosmology~\cite{2013Maldacena.PimentelJoHEP,2021Martin.VenninJoCaAP,2024Chandran.etalPRD,2025Belfiglio.etalPR}.

\begin{figure}[!t]
		\centering
	\includegraphics[scale=0.32]{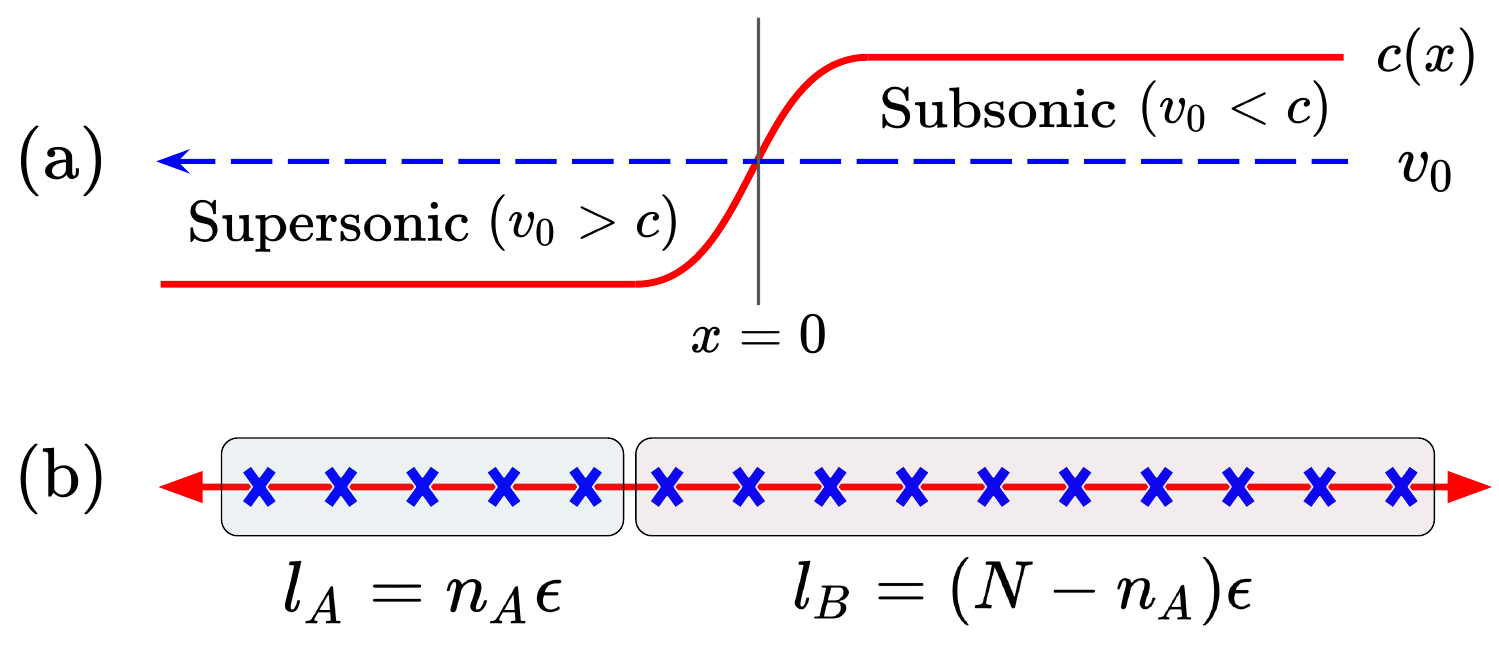}
				\caption{(a) Sound speed profile $c(x)$ that generates an analogue black-hole horizon at $x=0$. (b) Bipartition scheme for $N$ lattice points sampled along 
                $x\in\left[-
            \frac{L}{2},\frac{L}{2}\right]$.}
		\label{fig:illustration}
		\vspace*{-1em}
\end{figure}

\textit{Analogue black-hole setup.} We consider a 
BEC in the hydrodynamic regime, where low-energy phonon excitations propagate on an effective spacetime metric determined by the background-flow velocity $\vec{v_0}=-v_0\hat{\imath}$, the local speed of sound $c(x)$ and density $n(x)$~\cite{2008Balbinot.etalPRA,2008Carusotto.etalNJoP,2013Anderson.etalPRD}. In the Madelung representation $\Psi=\sqrt{n}e^{i\theta}$, the phase fluctuation $\theta_1$ on top of  the condensate ($\theta=\theta_0+\theta_1$ where $v_0 =\frac\hbar m\nabla\theta_0$ and $m$ is the atomic mass) 
obeys a Klein–Gordon equation in a curved geometry: 
\begin{equation}
    {\Box}{\theta}_1=0;\quad \Box=\frac{1}{\sqrt{-g}}\partial_{\mu}\left(\sqrt{-g}g^{\mu\nu}\partial_{\nu}\right),
\end{equation}
where the space-time is described by a PG metric:
\begin{equation}
    ds^2=g_{\mu\nu}dx^{\mu}dx^{\nu}=\frac{n}{mc}\left[-c^2dT^2+(\vec{dx}-\vec{v_0}dT)^2\right].
\end{equation}
For a sound-speed profile $c(x)$ as shown in \ref{fig:illustration}, a horizon forms at the point where $c=v_0$, separating the subsonic exterior from a supersonic interior. We note that dimensional reduction from 3+1D 
 to an effective 1+1D sonic spacetime for the quasi-1D BEC, rescaling the field as ${\theta}_1=\sqrt{mc/n\hbar l_{\perp}^2}{\theta}_1^{(2)}$, leads to 
an effective potential $V$ in the KG equation, $(\Box^{(\rm 2D)}+V)\theta_1^{(2)}=0$. For a quasi-1D BEC, the transverse trap width $l_{\perp}\ll\xi$, where $\xi=\frac\hbar{mc}$ is the healing length. In what follows, we omit the effective potential $V$ since it is not salient to a generic 2D black hole, and it encodes correlation peaks that are subleading to those from the Hawking effect~\cite{2024Anderson.etalCRP}.

\textit{Entanglement from tomography.} In 
the experiments, a primary observable is the equal-time 
density-density correlation $\langle n(x)n(x')\rangle$ measurable via in situ imaging of the condensate. This is however subject to finite spatial resolution, effectively 
sampling the field at a discrete set of points $\{j\}$, separated by a resolution scale $\epsilon \gtrsim  \xi$. Since the phonon field remains approximately Gaussian throughout the evolution, the two-point correlators pertaining to the sampled points fully describe the quantum state for the coarse-grained subsystem. 
By further partitioning these lattice points  
into adjacent blocks (see \ref{fig:illustration}), the entanglement between phonon subregions can be extracted via established techniques~\cite{2005Braunstein.LoockRMP,2005Ferraro.etal,2017Serafini}.


In phase space, Gaussian states are by definition 
characterized by a Wigner function of the form $W\propto e^{\vec{Q}^T\vec{\Sigma}\vec{Q}/2}$, 
where the $2N$ dimensional vector $\mathbf{Q}$ collects all the quadrature field variables as $Q_j=\tilde\varphi_j$,  $Q_{j+N}=\tilde\pi_j$ (which correspond to the phase and density fluctuations induced by phonons at the lattice points), and the covariance matrix $\mathbf{\Sigma}$ is defined as:
\begin{subequations}\label{eq:cov1}
\begin{align}
    &\mathbfs{\Sigma}=\begin{bmatrix}       \mathbfs{\Sigma}_{\varphi\varphi}&\mathbfs{\Sigma}_{\varphi\pi}\\\mathbfs{\Sigma}_{\varphi\pi}^T&\mathbfs{\Sigma}_{\pi\pi}    \end{bmatrix};\quad\quad\,\,\, (\Sigma_{\varphi\varphi})_{jk}=\frac{1}{2}\langle \{\hat{\tilde\varphi}_j,\hat {\tilde\varphi}_k\}\rangle,
        \\
        &(\Sigma_{\varphi\pi})_{jk}=\frac{1}{2}\langle \{\hat {\tilde\varphi}_j,\hat{\tilde\pi}_k\}\rangle
        ,\quad 
        (\Sigma_{\pi\pi})_{jk}=\frac{1}{2}\langle \{\hat{\tilde\pi}_j,\hat{\tilde\pi}_k\}\rangle,
    \end{align} 
\end{subequations}
where the tildes specify that the quadratures have been rescaled to ensure dimensionless entries in the covariance matrix. The commutation and uncertainty relations are hence compactly represented as follows: 
\begin{equation}
    [\hat{Q}_j,\hat{Q}_k]=i\Omega_{jk};\quad  \vec{\Sigma}+\frac{i}{2}\vec{\Omega}\geq \vec{0}; \quad \mathbf{\Omega}=\begin{bmatrix}
        O&\mathbb{I}\\-\mathbb{I}&O
    \end{bmatrix}\,,
\end{equation}
where $\mathbf{\Omega}$ is the symplectic matrix and $\mathbb{I}$ is the $N\times N$ identity matrix. Let us now partition the system into $A,B$-sectors as illustrated in \ref{fig:illustration}. A partial transposition with respect to $B$ invokes a 
time reversal of the subsystem modes --- which in the phase space corresponds to flipping the parity of momentum operators belonging to $B$. For the transposed covariance matrix $\bar{\vec{\Sigma}}$, 
the symplectic spectrum $\{\bar{\nu}_j\}$ can be obtained from the eigenvalues $\{\pm  \bar{\nu}_j\}$ of the matrix $i\mathbfs{\Omega}\bar{\mathbfs{\Sigma}}$ as a direct consequence of the Williamson theorem~\cite{1936WilliamsonAJoM}. The Partial Positive Transpose (PPT) criterion~\cite{1996PeresPRL,1997HorodeckiPLA,2000SimonPRL} asserts that $A$ and $B$ are entangled if the condition $\bar{\vec{\Sigma}}+\frac{i}{2}\vec{\Omega}\geq 0$ is violated, or equivalently, there is at least one eigenvalue such that $\bar{\nu}_j<1/2$. The resultant entanglement content can be quantified via \textit{logarithmic negativity} (LN), defined as below~\cite{2002Vidal.WernerPRA,2005PlenioPRL}:
\begin{equation}
    \mathcal{E}_N=-\sum_{j} \ln\left[\min\left(1,2\bar{\nu}_j\right)\right].
\end{equation}

\textit{Lattice regularization in the PG vacuum.} In an effective 2D background described by the PG metric, the action for a massless scalar field takes the form~\cite{1982Birrell.Davies}:
\begin{equation}
    S=\int \frac{dTdx}{2c}
\left[\dot{\varphi}^2-2v_0\dot{\varphi}\partial_x\varphi-(c^2-v_0^2)\left(\partial_x\varphi\right)^2\right],
\end{equation}
where we set $\hbar=m=1$, and the field $\varphi =\theta_1^{(2)}$ and momentum $ \pi=(\dot{\varphi}-v_0\partial_x\varphi)/c= -(cl_{\perp}^2/n)^{1/2}n_1$ map respectively to the phase and density fluctuations. The PG vacuum correlators are obtained as:
\begin{align}\label{eq:gtransport}
    \langle\{\hat{\varphi}(x,T),\hat{\varphi}(x',T')\}\rangle&=\int_{\omega_0}^{\omega_1}\!\!\frac{d\omega}{2\pi\omega}\cos{\left(\frac{\omega A}{2}\right)}\cos{\left(\frac{\omega B}{2}\right)}\nonumber\\\langle\{\hat{\varphi}(x,T),\hat{\pi}(x',T')\}\rangle&=\frac{1}{c}D_{T',x'} \langle\{\hat{\varphi}(x,T),\hat{\varphi}(x',T')\}\rangle\nonumber\\
    \langle\{\hat{\pi}(x,T),\hat{\pi}(x',T')\}\rangle&=\frac{1}{c}D_{T,x}\langle\{\hat{\varphi}(x,T),\hat{\pi}(x',T')\}\rangle\! 
\end{align}
where $D_{T,x}=\partial_{T}-v_0\partial_{x}$, $A=u-u'-v+v'$, and $B=u-u'+v-v'$, in terms of retarded/advanced null coordinates $u=T-\int \frac{dx}{c-v_0}$ and $v=T+\int \frac{dx}{c+v_0}$. The IR ($\omega_0$) and UV ($\omega_1$) cutoffs prevent the equal-time correlators ($T=T'$) from diverging in the coincidence limit $x\to x'$.

For the extraction of finite entanglement measures from spatial bipartitions of the quantum field in a black-hole background, we now present a framework that goes beyond standard regularization techniques~\cite{1986Bombelli.etalPRD,1993SrednickiPRL,1997Mukohyama.etalPRD}. 
Conventionally, a \textit{harmonic lattice} prescription is adopted wherein the UV-regulator $\epsilon$ is set by the lattice spacing between the discretized degrees of freedom $\varphi_j=\varphi(x=j\epsilon)$, and the IR-regulator is set by the boundary at $L=(N+1)\epsilon$. A 
discrete Hamiltonian is then derived corresponding to $N$ coupled harmonic oscillators i.e., $H=\frac{1}{2}\sum_j(\pi_j^2+\sum_{k}K_{jk}\varphi_j\varphi_k)$, 
which allows us to write down the wavefunction and extract the leading order entanglement from simply a finite number of modes. The main advantage of this approach is that although the measured entanglement is UV-sensitive, the scaling law as such is robust to the cutoff choice. 
However, it does not have a straightforward extension to the black-hole interior
(the corresponding degrees of freedom 
must be walled off to have a positive semidefinite
$K_{ij}$~\cite{1998Mukohyama.etalPRD,2020Chandran.ShankaranarayananPRD}), whereas the harmonic lattice fails to capture the coarse-graining inherent to experimental detection. To address these issues, we propose a 
regularization scheme implemented at the level of the covariance matrix~\eqref{eq:cov1}, incorporating the spatial resolution limits of correlation measurements pertaining to $N$ lattice points sampled from the continuum, along $x_j=-\frac{L}{2}+j\epsilon\in\left[-
            \frac{L}{2},\frac{L}{2}\right]$ (\ref{fig:illustration}).

\begin{figure}[!tb]
\vspace*{1em}
		\centering
	\includegraphics[scale=0.45]{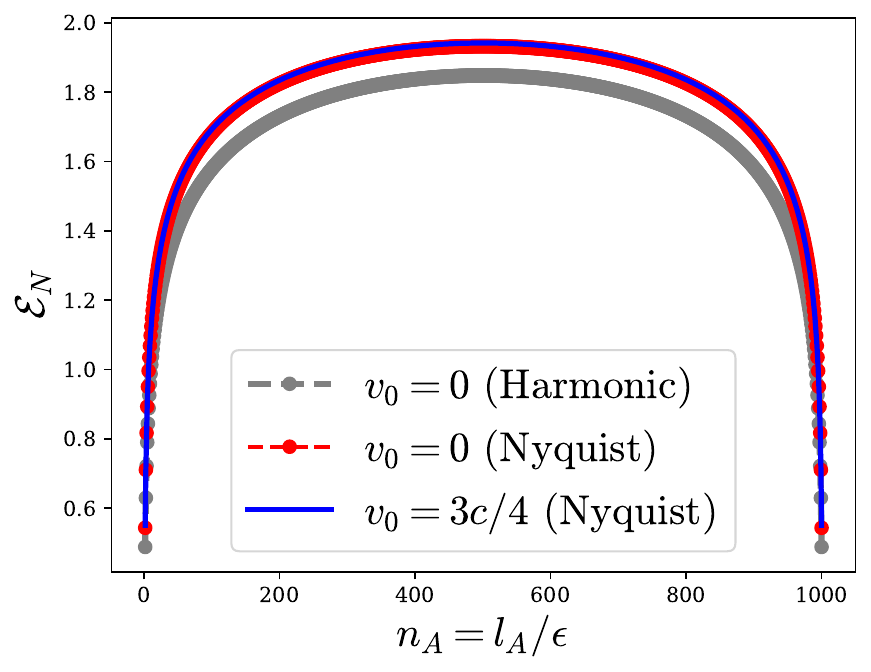}
				\caption{Vacuum scaling of logarithmic negativity $\mathcal{E}_N$ corresponding to 
				Nyquist and harmonic lattices of size $N=10^3$, which differ by an 
                an additive constant.}
		\label{fig:VacuumNeg}
\end{figure}

For the conformal PG vacuum corresponding to a constant, homogeneous sound speed $c$, the covariance matrix can be decomposed into outgoing ($\vec{I}$) and incoming ($\vec{J}$) branches of the mode functions as $\vec{\Sigma}_{\varphi\varphi}=\vec{I}_{\varphi\varphi}+\vec{J}_{\varphi\varphi}$. The coarse-grained elements are then given by:
\begin{subequations}
\begin{align}  
(I_{\varphi\varphi})_{jk}&=\int_{\tilde{\omega}_{0u}}^{\tilde{\omega}_{1u}}\frac{d\tilde{\omega}}{4\pi\tilde{\omega}}\cos{\left(\frac{\tilde{\omega}|j-k|}{1-\tilde{v}_0}\right)}\\
(J_{\varphi\varphi})_{jk}&=\int_{\tilde{\omega}_{0v}}^{\tilde{\omega}_{1v}}\frac{d\tilde{\omega}}{4\pi\tilde{\omega}}\cos{\left(\frac{\tilde{\omega}|j-k|}{1+\tilde{v}_0}\right)}\,\,,
\end{align}
\end{subequations}
where $\tilde{v}_0=v_0/c$ and $\tilde{\omega}=\omega \epsilon/c$ are dimensionless parameters. 
To fix the integration limits, we first invoke the result that the leading-order term of negativity is independent of the IR cutoff, but UV-sensitive due to divergent short-distance correlations~\cite{2012Calabrese.etalPRL}. We may therefore set the IR cutoff arbitrarily such that $\tilde{\omega}_{0u},\tilde{\omega}_{0v}\ll1$, far below any relevant low-energy scales. 
For the UV cutoff we employ a Nyquist lattice~\cite{2015Pye.etalPRD,2023Lewis.etalPRD} corresponding to $\rm \epsilon_{Nyq}\equiv \pi/k_{\rm UV}$, where $k_{\rm UV}$ fixes the \textit{bandwidth} of correlation measurements \footnote{Rather than discretizing bandlimited fields as done in~\cite{2015Pye.etalPRD,2023Lewis.etalPRD}, we simply coarse-grain bandlimited correlations (the primary observables here) from the continuum. See Sec.\,\,II.B of \cite{SM} for a detailed discussion on our approach, including non-Nyquist choices.}. Such a lattice admits a faithful reconstruction of the coarse-grained state relative to the allowed bandwidth, where the short-distance correlations are regularized by the resolution limits of the measurement. In the frequency domain, this \textit{Nyquist choice} is incorporated via dispersion relations for the outgoing and incoming UV modes, yielding distinct cutoffs for the respective branches of the two-point correlators:
\begin{equation}\label{eq:w-cutoff}
\tilde{\omega}_{1u}=(1-\tilde{v}_0)\pi;\quad \tilde{\omega}_{1v}=(1+\tilde{v}_0)\pi.
\end{equation}
Having fixed these cutoffs for the covariance matrix elements, we validate our approach by extracting the negativity scaling for adjacent intervals ($l_A+l_B=L$, see \ref{fig:VacuumNeg}) that exactly matches the 
known CFT result~\cite{2012Calabrese.etalPRL} 
\begin{equation}\label{eq:vacscaling}
    \mathcal{E}_N=\frac{1}{4}\ln{\left[\frac{l_Al_B}{L\epsilon}\right]}+\rm const.\,,
\end{equation}
up to a nonuniversal additive constant. 
Our prescription~\eqref{eq:w-cutoff} 
hence ensures that the coarse-graining preserves 
the universal scaling law of vacuum entanglement down to the prefactor, from simply a finite sample of correlators. The scaling law is robust to taking the continuum limit of $\epsilon\to0$ and $N\to\infty$, where the characteristic UV-divergence of entanglement emerges. 
We now extend our approach to an analogue black-hole background. 

\textit{Entanglement scaling of HR.} For a sound-speed profile that simulates a black-hole horizon at $x=0$~(\ref{fig:illustration}), we assume that the {\em surface gravity} $\kappa=dc/dx|_{x=0}$ remains sub-Planckian ($\kappa \ll c\xi^{-1}$) to safely stay in a regime where sonic Lorentz invariance holds. We consider the Unruh state~\cite{1976UnruhPRD}, which yields an outgoing thermal flux of Hawking quanta from the horizon --- in the condensate, it appropriately captures the late-time quasiparticle 
correlations generated by a black-hole flow~\cite{2008Balbinot.etalPRA,2008Carusotto.etalNJoP,2013Anderson.etalPRD,2021Fabbri.BalbinotPRL}. 
Upon neglecting the effective potential, the mode functions reduce to simple plane-wave solutions, resulting in the following form for the two-point correlators along the interior (L) and exterior (R) regions (see Sec.\,\,I.B of \cite{SM}),
\begin{subequations}
\begin{align}
(I_{\varphi\varphi})_{jk}&=\int_{\tilde{\omega}_{0u}}^{\tilde{\omega}_{1u}}\frac{d\tilde{\omega}}{4\pi\tilde{\omega}}\,f_\omega(\tilde{\kappa})\cos{\left[\tilde{\omega}(\tilde{u}_j-\tilde{u}_k)\right]}\\ (J_{\varphi\varphi})_{jk}&=\int_{\tilde{\omega}_{0v}}^{\tilde{\omega}_{1v}}\frac{d\tilde{\omega}}{4\pi\tilde{\omega}}\,\cos{\left[\tilde{\omega}(\tilde{v}_j-\tilde{v}_k)\right]}\\
        f_{\omega}(\tilde{\kappa})&=\begin{cases}
        \coth\left(\frac{\pi\tilde{\omega}}{\tilde{\kappa}}\right) & \text{for L-L or R-R} \\
        \csch\left(\frac{\pi\tilde{\omega}}{\tilde{\kappa}}\right) &  \text{for L-R}
    \end{cases}\,\,\,\,\,\,,
\end{align}
\end{subequations}
where $c_{L,R}=c(\mp\infty)$, $\tilde{\kappa}=\frac{\epsilon}{c_R}\kappa $, $\tilde{\omega}=\frac{\epsilon}{c_R}\omega$, $\tilde{u}_j=\frac{c_R}{\epsilon} u(x_j)$ and $\tilde{v}_j=\frac{c_R}{\epsilon} v(x_j)$. The other correlators are derived via Gaussian transport equations given in \eqref{eq:gtransport}. For L-R correlations, a positive peak emerges in the $\vec{I_{\varphi\varphi}}$ component when $\tilde{u}_j\to \tilde{u}_k$, i.e., along the paths of Hawking quanta:
\begin{equation}\label{eq:corrpeak}
    \frac{x'}{c_L-v_0}=\frac{x}{c_R-v_0}\,\,,
\end{equation}
  where $x'<0$ and $x>0$. For the $\vec{I_{\pi\pi}}$ component, and in turn, the density-density correlations, these points correspond to a \textit{negative} peak 
  of the order $\mathcal{O}(\kappa^2)$, that serves as a reliable signature of the Hawking effect~\cite{2008Balbinot.etalPRA,2008Carusotto.etalNJoP,2013Anderson.etalPRD}. While these peaks have subsequently been observed in the laboratory~\cite{2016SteinhauerNP,2019MunozdeNova.etalN,2021Kolobov.etalNP}, additional subleading peaks that were predicted (associated with the effective potential) are yet to be experimentally confirmed. 

\begin{figure}[tb]
	\begin{center}
		\subfloat[][$v_0/c_R=3/4$]{%
			\includegraphics[width=0.24\textwidth]{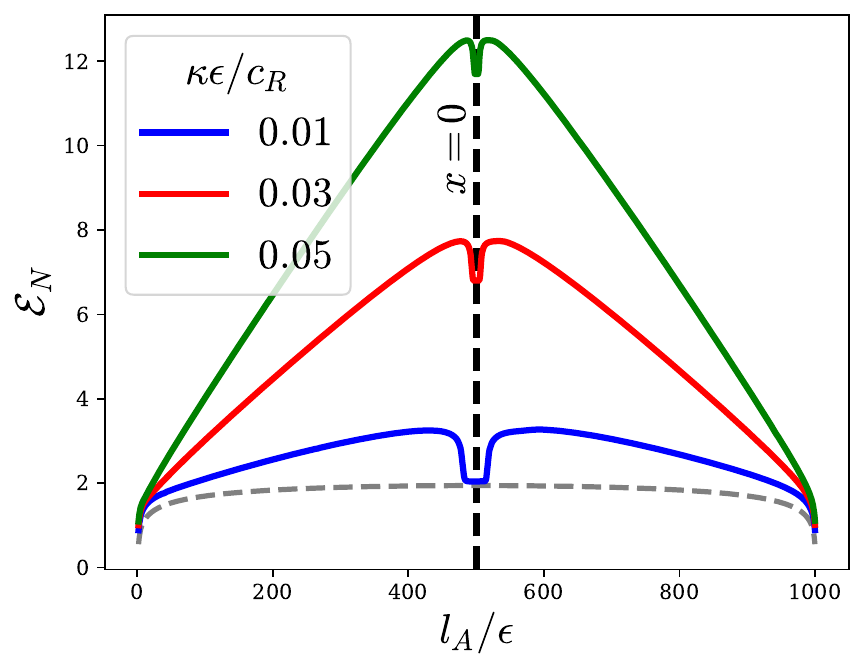}\label{fig:HR1a}
		}
		\subfloat[][$\kappa\epsilon/c_R=0.1$]{%
			\includegraphics[width=0.24\textwidth]{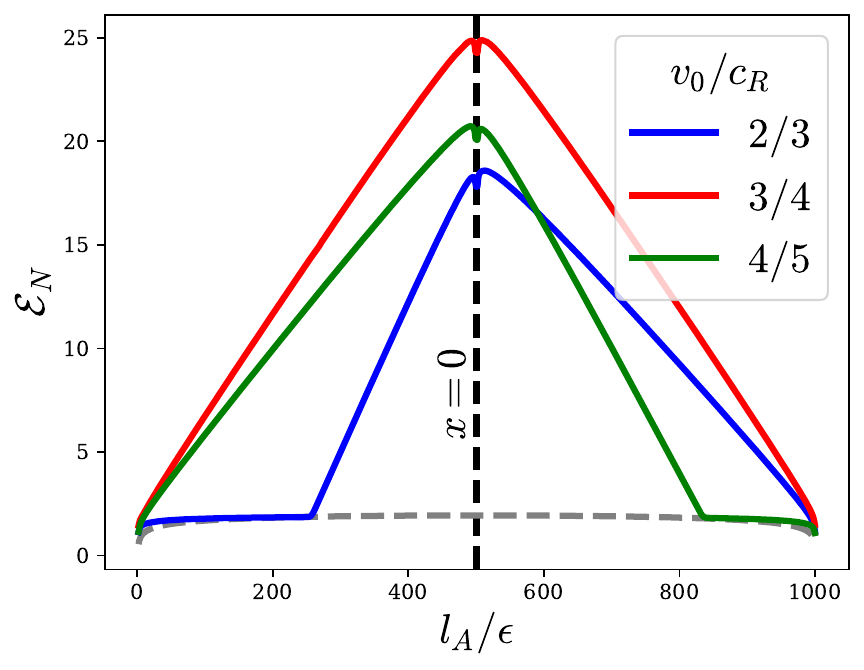}\label{fig:HR1b}
		} 
		\caption{Logarithmic negativity scaling with subsystem size $n_A$ for $N=10^3$, $c_L/c_R=1/2$, and various values of (a) the surface gravity $\kappa$ and (b) flow velocity $v_0$. The dashed gray line corresponds to the PG vacuum scaling, and the dashed black line locates the analogue horizon.}
		\label{fig:HR1}
	\end{center}
	\vspace*{-1em}
\end{figure}

Extending our regularization scheme~\eqref{eq:w-cutoff} to black-hole subregions, we fix the UV cutoffs as $\tilde{\omega}_{1u}^L=(\tilde{v}_0-\tilde{c}_L)\pi$, $\tilde{\omega}_{1v}^L=(\tilde{v}_0+\tilde{c}_L)\pi$ for L-L correlations, and $\tilde{\omega}_{1u}^R=(1-\tilde{v}_0)\pi$, $\tilde{\omega}_{1v}^R=(1+\tilde{v}_0)\pi$ for R-R correlations --- where we have defined $\tilde{c}_L=c_L/c_R$ and $\tilde{v}_0=v_0/c_R$. For L-R correlations, we take the harmonic mean of the above frequencies, i.e., $\tilde{\omega}_{1u}=2/(1/\tilde{\omega}_{1u}^L+1/\tilde{\omega}_{1u}^R)$ and $\tilde{\omega}_{1v}=2/(1/\tilde{\omega}_{1v}^L+1/\tilde{\omega}_{1v}^R)$ --- however we note that the results presented here are insensitive to this choice, as the coincidence limit is avoided in strictly L-R correlations. We consider the range $0.01\leq\tilde{\kappa}\leq0.1$ to stay sufficiently close to the hydrodynamical 
regime (setting $\epsilon \sim\xi$)~\cite{2008Carusotto.etalNJoP}. Following the same bipartition scheme as for the vacuum case, we see from \ref{fig:HR1} that a volume term emerges for the Unruh state as a direct consequence of the nonlocal correlations seeded by HR. As visible from \ref{fig:HR1}\subref{fig:HR1a}, this scaling is only resolvable beyond the quantum atmosphere~\cite{2016GiddingsPLB,2017Dey.etalPLB} --- the near-horizon region up to a distance $\sim \mathcal{O}(v_0/\kappa)$ (located  
in and around the central dips) 
where the nonlocal peaks are 
dominated by short-distance correlations~\cite{2021Fabbri.BalbinotPRL}. 

\begin{figure}[hbt]
	\begin{center}
		\subfloat[$l_A<l_H$]{%
			\includegraphics[width=0.24\textwidth]{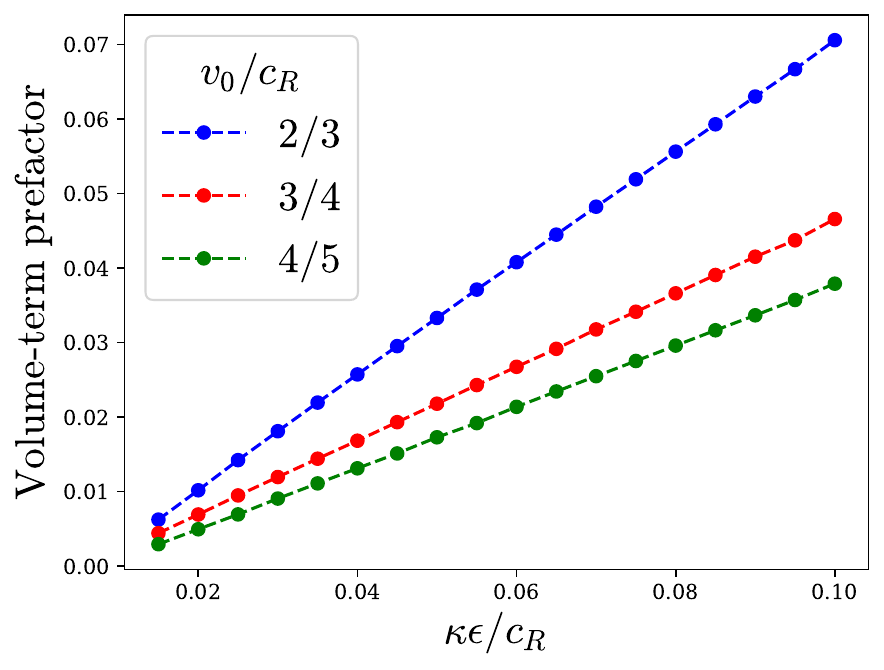}\label{fig:HR2a}
		}
		\subfloat[$l_A>l_H$]{%
			\includegraphics[width=0.24\textwidth]{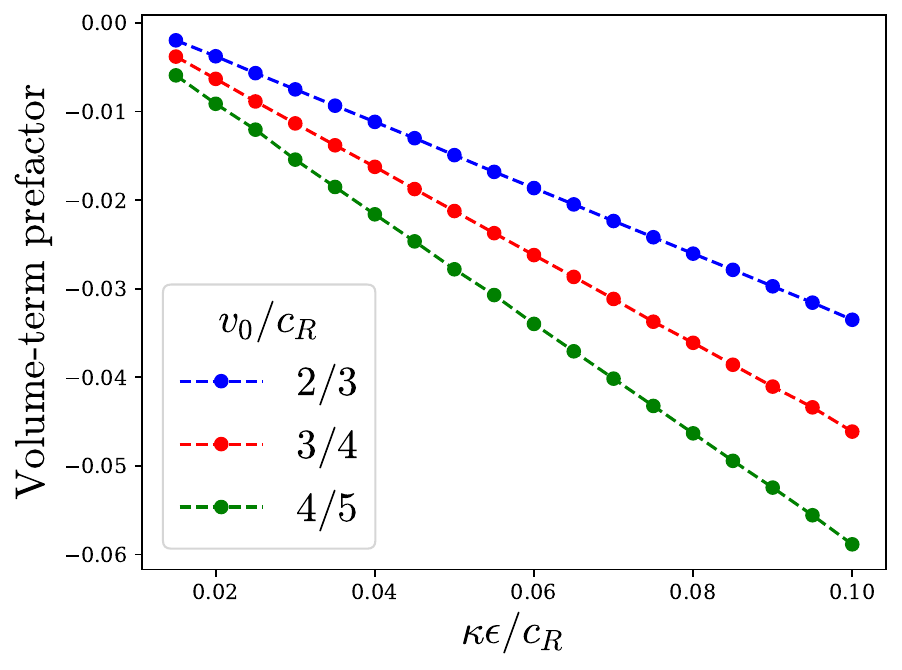}\label{fig:HR2b}
		}
		\caption{Prefactor of negativity volume-term ($\propto n_A$): Dependence on surface gravity $\kappa$ for (a) the black-hole interior and (b) exterior. Here, $N=600$ and $\tilde{c}_R=1/2$.}
		\label{fig:HR2}
	\end{center}
	\vspace*{-1em}
\end{figure}

Our main results are as follows. First, from \ref{fig:HR2} we see that the slope scales linearly with $\tilde{\kappa}$, due to which the regulator $\epsilon$ cancels out ($\tilde{\kappa}n_A=\kappa l_A/c_R$) and the volume term is revealed to be \textit{UV-finite}. The overall scaling can thus be decomposed into a UV-sensitive term that captures the regulated short-distance correlations, and a UV-finite volume term corresponding to nonlocal correlations generated by HR, i.e., as $\mathcal{E}_N\sim \mathcal{E}_N^{\rm (UV)}+\mathcal{E}_N^{\rm (HR)}$. Second, we know from \eqref{eq:corrpeak} that the nonlocal peaks are distributed asymmetrically about the horizon for unequal mode velocities in the interior and exterior regions. This is reflected in \ref{fig:HR1}\subref{fig:HR1b} (the blue and green lines) via asymmetric slopes for the 
volume term, and a recurrence of the vacuum scaling away from the horizon. The latter simply indicates that the bipartition fails to capture both members of the entangled Hawking pair (now asymmetrically distributed) and therefore does not contribute to  negativity. 
From linear fits of \ref{fig:HR2}, we numerically fix the 
negativity scaling of HR 
as (see Sec.\,\,II.C of \cite{SM}):
\begin{equation}
        \mathcal{E}_N^{\rm (HR)} (l_A) \sim \frac{\kappa}{8}\left[\frac{l_H}{v_{\rm{H}}^{\rm \max}} -\frac{|l_A-l_H|}{v_{\rm H}(l_A)}\right],
\end{equation}    
where $v_{\rm H}$ is the velocity of Hawking quanta in the interior ($v_{\rm H}|_{l_A<l_H}=v_0-c_L$) or the exterior ($v_{\rm H}|_{l_A>l_H}=c_R-v_0$) regions, $v_{\rm H}^{\rm max}=\max[v_0-c_L,c_R-v_0]$ is the larger of the two, and $l_H$ is the subsystem size up to the horizon at $x=0$. 
For equal velocities on either side, we in fact get $\mathcal{E}_N^{\rm (HR)}\sim \frac{\kappa}{8v_{\rm H}}\min{\left[l_A,l_B\right]}$, exactly resembling the Page curve behavior up to a prefactor~\cite{2021Shapourian.etalPQ}. The scaling always peaks near the horizon (outside of the quantum atmosphere) regardless of how the lattice points are sampled for the bipartition, 
and the entanglement is 
strongly amplified relative to the vacuum state (which we also expect for thermal states~\cite{SM}). 
The $\mathcal{O}(\kappa^2)$ correlation peak is elevated to a more prominent $\mathcal{O}(\kappa)$ term in the negativity scaling, growing linearly with both the number density of Hawking pairs ($n\propto \kappa/v_{\rm H}$) and the probed spatial region ($\propto l_H$). The scaling is therefore 
a 
sensitive signature of HR pertaining to the entanglement content as well as the \textit{spatial distribution} of Hawking pairs along the black-hole interior/exterior.  Furthermore, it is independent of the UV-regulator $\epsilon$, serving as a robust information-theoretic imprint of the nonlocal density peaks that have already been observed in analogue experiments~\cite{2016SteinhauerNP,2019MunozdeNova.etalN,2021Kolobov.etalNP}.

\textit{Discussion.} We have developed a novel lattice-regularization scheme for the extraction of entanglement measures from equal-time  
correlators, readily adaptable to 
theoretical and experimental approaches to
curved space-time QFT. Employing this, we have resolved for the first time the geometric scaling of entanglement arising from Hawking radiation in an experimentally accessible analogue black-hole model. Our central finding is that logarithmic negativity --- which typically exhibits a UV-divergent log-law pertaining to the universal short-distance correlation structure of the vacuum --- acquires a UV-finite volume term arising from the nonlocal correlations generated by HR. The volume-term prefactor is sensitive to the surface gravity as well as the outgoing mode velocities. It therefore records information about both the number density and spatial distribution of entangled Hawking pairs across the black-hole horizon, serving as a genuine entanglement witness of HR that is robust to how short-distance correlations are regularized.


Within the quantum atmosphere, short-distance correlations dominate over Hawking correlations, suppressing the volume-law contribution. Notably, this contribution is also absent in earlier studies of black-hole entanglement which focused primarily on generic quantum states states~\cite{1986Bombelli.etalPRD,1993SrednickiPRL,2006Das.ShankaranarayananPRD,2007Das.ShankaranarayananCaQG}, or invoked restrictive assumptions on the lattice 
--- such as in ``brick-wall" constructions~\cite{1997Mukohyama.etalPRD,1998Mukohyama.IsraelPRD,2020Chandran.ShankaranarayananPRD,2025Belfiglio.etalPRD} where the interior degrees of freedom are 
walled off and the bipartition is imposed strictly outside the horizon. 
 The resultant area-law scaling obtained 
in the exterior should still hold for near-horizon partitions in the low-$\kappa$ regime (large black holes), where the nonlocal peaks from HR are negligible and the quantum atmosphere is extended. For smaller (hotter) black holes, however, the interior and exterior modes are strongly entangled, 
leading to an enhanced volume-law contribution. Our results in fact show how HR can trigger a Page-curve-like scaling ~\cite{1993PagePRL,2022Bianchi.etalPQ,2021Shapourian.etalPQ} in the early stages of evaporation (well before the Page time~\cite{2013PageJoCaAP}).  We thereby establish pair creation as an underlying mechanism for the emergence of volume-law entanglement, consistent with recent results for the Sauter-Schwinger effect~\cite{2026Chandran.Rajeev}. 


Our findings have immediate implications:  
First, the volume term predicted here should be observable in currently realizable analogue black-hole experiments~\cite{2016SteinhauerNP,2019MunozdeNova.etalN,2021Kolobov.etalNP}. The incorporation of mode backscattering via the effective potential~\cite{2024Anderson.etalCRP,2024Anderson.etalPRD} dispersive effects at high momenta~\cite{2005Balbinot.etalRNC,2023HolandaRibeiro.FischerPRD}, and nonlinear effects from quantum backreaction~\cite{2005Schuetzhold.etalPRD,2024Pal.FischerPRD} can further elucidate the fine structure of entanglement scaling. Second, the regularization framework provides a broader blueprint for extracting finite entanglement measures for not only stationary black-hole backgrounds~\cite{2020Chandran.ShankaranarayananPRD,2025Belfiglio.etalPRD}, but also dynamical models involving cosmological expansion~\cite{2024Chandran.etalPRD,2025Chandran.FischerTEPJC} and gravitational collapse~\cite{2022Balbinot.FabbriPRD,2025Belfiglio.etalPRDa,2025Sahota.etalPRD}. These represent the natural next steps toward probing the emergence of Page curve~\cite{1993PagePRLa,2013PageJoCaAP} from horizon bipartitions in black-hole evaporation models.

This work was supported by the 
NRF of Korea under 
Grant No.~2020R1A2C2008103. It has not been
supported by the 
IRC NextQuantum at Seoul National University.



\bibliography{HRNeg_v8}


\newpage
\begin{widetext}
\setcounter{equation}{0}
\setcounter{figure}{0}
\setcounter{table}{0}
\setcounter{page}{1}
\renewcommand{\theequation}{S\arabic{equation}}
\renewcommand{\thefigure}{S\arabic{figure}}

\section*{Supplemental Material}

\subsection{I. Analogue black-hole review}
The Bogoliubov-de Gennes equations in the hydrodynamical approximation are given by:
\begin{equation}
   \partial_T{\theta}_1=- \vec{v_0}\vec{\nabla}{\theta}_1-\frac{mc^2}{n\hbar}{n}_1\quad;\quad \partial_T {n}_1=-\vec{\nabla}\left(\vec{v_0}{n}_1+\frac{\hbar n}{m}\vec{\nabla}{\theta}_1\right) \,,
\end{equation}
where $\theta_1$ and $n_1$ are the phase and density fluctuations respectively, the flow $\vec{v_0}=\hbar\nabla\vec{\theta}_0/m=-v_0\hat{\imath}$ is uniform and stationary, and the sound speed $c=\sqrt{ng/m}$ is generally inhomogeneous. Dimensionally reducing to 2D, we get~\cite{2013Anderson.etalPRD}:
\begin{equation}
    (\Box^{(2)}+V)\theta_1^{(2)}=0;\quad {\theta}_1=\sqrt{\frac{mc}{n\hbar l_{\perp}^2}}{\theta}_1^{(2)};\quad \Box^{(2)}=-\frac{1}{c^2}\partial_T^2+\frac{1}{c}\partial_x(c\partial_x)-\frac{2v_0}{c^2}\partial_T\partial_x
\end{equation}
where the transverse trap width is given by $l_{\perp}\ll\xi$ ($\xi\equiv\hbar/mc$ is the healing length), and the effective potential:
\begin{equation}
    V=\frac{1}{2c}\frac{d^2c}{dx^2}\left(1-\frac{v_0^2}{c^2}\right)-\frac{1}{4c^2}\left(\frac{dc}{dx}\right)^2+\frac{5v_0^2}{4c^4}\left(\frac{dc}{dx}\right)^2
\end{equation}
Note that we have rescaled $\Box^{(2)}$ and $V$ to have the usual dimensions of $l^{-2}$, as opposed to their forms in \cite{2013Anderson.etalPRD}. In what follows, we ignore the effective potential $V$ as it is a higher-dimensional curvature effect that is not salient to generic 2D spacetimes. Furthermore, its effects are subleading to the Hawking radiation signatures we discuss in this work.  

\subsubsection{A. Vacuum correlators}
Setting $\hbar=m=1$, the effective 2D metric and the corresponding action for a massless scalar field $\varphi$ is given by:
\begin{equation}
    ds^2=-c^2(x)dT^2+(dx+v_0dT)^2\quad;\quad  S=\frac{1}{2}\int \frac{dTdx}{c(x)}
\left[\dot{\varphi}^2-2v_0\dot{\varphi}\partial_x\varphi-\left\{c^2(x)-v_0^2\right\}\left(\partial_x\varphi\right)^2\right],
\end{equation}
where $\varphi\equiv\theta_1^{(2)}$. The conjugate momentum is given by $ \pi=\partial L/\partial \dot{\varphi}=(\dot{\varphi}-v_0\partial_x\varphi)/c$. This is related to the density fluctuation as $\pi\equiv \frac{v_0}{2c^2}\varphi\partial_xc-(cl_{\perp}^2/n)^{1/2}n_1$ --- however the first term here can be absorbed into the effective potential $V$ via a canonical transformation. For constant $c$ (or away from the modulation region for black-hole flow), we get:
\begin{equation}
    \varphi=\sqrt{\frac{nl_{\perp}^2}{c}}\theta_1\quad;\quad \pi=-\sqrt{\frac{cl_{\perp}^2}{n}}n_1
\end{equation}
We now promote the variables to operators that satisfy commutation relations $[\hat{\varphi}(T,x),\hat{\pi}(T,x')]=i\delta(x-x')$.  For the PG vacuum, defined as $\hat{a}_{u,\omega}|0\rangle=\hat{a}_{v,\omega}|0\rangle=0$ $\forall\omega>0$, the quantized field can be expanded as:
\begin{equation}
    \hat{\varphi}=\int_0^{\infty} d\omega \left[\hat{a}_{u}(\omega)\frac{e^{-i\omega u}}{\sqrt{4\pi\omega}}+\hat{a}_{u}^{\dagger}(\omega)\frac{e^{i\omega u}}{\sqrt{4\pi\omega}}+\hat{a}_{v}(\omega)\frac{e^{-i\omega v}}{\sqrt{4\pi\omega}}+\hat{a}_{v}^{\dagger}(\omega)\frac{e^{i\omega v}}{\sqrt{4\pi\omega}}\right]\,,
\end{equation}
in terms of the retarded and advanced null coordinates defined respectively as $u=T-\int \frac{dx}{c-v_0}$ and $v=T+\int \frac{dx}{c+v_0}$, and where the commutation relations $[\hat{a}_{u}(\omega),\hat{a}_{u}(\omega')^\dagger]=[\hat{a}_{v}(\omega),\hat{a}_{v}(\omega')^\dagger]= \delta(\omega-\omega')$ are satisfied. The vacuum state correlators are therefore obtained as follows:
\begin{subequations}\label{eq:vacuumcorrs}
    \begin{align}
        \langle \{\hat{\varphi}(x,T),\hat{\varphi}(x',T')\}\rangle&=\int\frac{d\omega}{2\pi\omega}\left[\cos{[\omega(u-u')]}+\cos{[\omega(v-v')]}\right]\\
        \langle \{\hat{\varphi}(x,T),\hat{\pi}(x',T')\}\rangle&=\int\frac{d\omega}{2\pi}\left[\frac{\sin{[\omega(u-u')]}}{c-v_0}+\frac{\sin{[\omega(v-v')]}}{c+v_0}\right]\\
        \langle \{\hat{\pi}(x,T),\hat{\pi}(x',T')\}\rangle&=\int\frac{d\omega\,\omega}{2\pi}\left[\frac{\cos{[\omega(u-u')]}}{\left(c-v_0\right)^2}+\frac{\cos{[\omega(v-v')]}}{\left(c+v_0\right)^2}\right],
    \end{align}
\end{subequations}
which can be split into incoming ($\vec{J}$) and outgoing ($\vec{I}$) contributions as discussed in the main text.

\subsubsection{B. Correlators for the Unruh state}

We consider the following sound speed profile for simulating a black-hole horizon at $x=0$~\cite{2013Anderson.etalPRD}:
\begin{align}
    c(x)&=\sqrt{c_L^2+\frac{(c_R^2-c_L^2)}{2}\left[1+\frac{2}{\pi}\tan^{-1}\left(\frac{x+b}{\sigma_v}\right)\right]}\\
    \sigma_v&=\frac{c_R^2-c_L^2}{2\pi v_0\kappa}\sin^2\left[\frac{\pi(v_0^2-c_R^2)}{c_R^2 - c_L^2}\right]\\
    b&=\sigma_v\tan\left[\frac{\pi(2v_0^2-(c_L^2+c_R^2))}{2(c_R^2-c_L^2)}\right]\,,
\end{align}
where the \textit{surface gravity} $\kappa\equiv\frac{dc}{dx}|_{x=0}$. For the above profile, numerical results show excellent quantitative agreement with the gravitational prediction in the hydrodynamical limit $\sigma_v \gg \xi$~\cite{2008Carusotto.etalNJoP}. We therefore consider the range of surface gravity values $0.01\leq\tilde{\kappa}\equiv\frac{\kappa\epsilon}{c_R}\lesssim 0.1$ in order to cover this regime (where $\epsilon\sim\xi$). We urge the readers to follow \cite{2013Anderson.etalPRD} for the complete calculation of correlators for the Unruh state. Here, we briefly review the calculation for the special case of $V=0$, i.e., ignoring the backscattering effects arising from the effective potential~\cite{2008Balbinot.etalPRA}.

Shifting to Schwarzschild-like time ($t$) and Tortoise coordinate ($x^*$), and neglecting the effective potential, we get:
\begin{equation}
    ds^2=-[c^2(x)-v_0^2](dt^2-dx_*^2)\quad \implies \quad \left(\partial_t^2-\partial_{x^*}^2\right)\varphi=0
\end{equation}
where the coordinates are in defined in left ($x<0$) and right ($x>0$) regions as follows:
\begin{equation}\label{eq:coord1}
    t=\begin{cases}
        T-\int_{X_2}^x dy \frac{v_0}{c^2(y)-v_0^2}+a & x<0 \\
        T-\int_{X_1}^x dy \frac{v_0}{c^2(y)-v_0^2} &  x>0
    \end{cases}\quad ;\quad x^*=\begin{cases}
        \int_{X_4}^x dy\frac{c(y)}{c^2(y)-v_0^2} & x<0 \\
       \int_{X_3}^x dy\frac{c(y)}{c^2(y)-v_0^2} &  x>0
    \end{cases}\quad,
\end{equation}
where $X_1$, $X_2$, $X_3$, and $X_4$ are constants that we fix later. The retarded ($u$) and advanced ($v$) null coordinates now take the form $v=t+x^*$ and $u=t-x^*$. The continuity of $v$ across the horizon allows us to fix $a$ as follows:
\begin{equation}
    a=\int_{X_1}^{X_2} \frac{dy}{c(y)+v_0}+\int_{X_3}^{X_1} dy\frac{c(y)}{c^2(y)-v_0^2} +\int_{X_2}^{X_4} dy \frac{c(y)}{c^2(y)-v_0^2}
\end{equation}
For the next steps, we also introduce the Kruskal coordinates:
\begin{equation}\label{eq:Kruskal}
    U_K=\begin{cases}
        e^{-\kappa u}/\kappa  & x<0 \\
        -e^{-\kappa u}/\kappa &  x>0
    \end{cases}\quad ;\quad V_K=\begin{cases}
        e^{\kappa v}/\kappa & x<0 \\
       e^{\kappa v}/\kappa &  x>0
    \end{cases}\quad,
\end{equation}
The Unruh state is defined such that the retarded modes originating from the horizon (at past infinity) are positive frequency with respect to the Kruskal coordinate $U_K$ (i.e., $\sim e^{-i\omega U_K}$). However in the stationary background, this would correspond to an outgoing flux of thermal radiation from the past horizon. There is also no incoming flux coming from past null infinity --- the advanced modes therefore have a similar form as the PG conformal vacuum ($\sim e^{-i\omega v}$). The quantized field for the Unruh state can therefore expanded as follows:
\begin{equation}
    \hat{\varphi}=\int_0^{\infty} d\omega_K \left[\hat{a}_{K}(\omega_K)\frac{e^{-i\omega_K U_K}}{\sqrt{4\pi\omega_K}}+\hat{a}_{K}^{\dagger}(\omega_K)\frac{e^{i\omega_K U_K}}{\sqrt{4\pi\omega_K}}\right]+\int_0^{\infty} d\omega \left[\hat{a}_{v}(\omega)\frac{e^{-i\omega v}}{\sqrt{4\pi\omega}}+\hat{a}_{v}^{\dagger}(\omega)\frac{e^{i\omega v}}{\sqrt{4\pi\omega}}\right]
\end{equation}
We can then expand the positive frequency modes (coming from the past horizon) in terms of the retarded modes in the interior (L) and exterior (R) regions as follows:
\begin{equation}
    \frac{e^{-i\omega_KU_K}}{\sqrt{4\pi\omega_K}}=
        \int_0^{\infty} d\omega \left[\left(\alpha^{L}_{\omega,\omega_K}\frac{e^{i\omega u}}{\sqrt{4\pi\omega}}+\beta^{L}_{\omega,\omega_K}\frac{e^{-i\omega u}}{\sqrt{4\pi\omega}}\right) \Theta(-x)+ 
        \left(\alpha^{R}_{\omega,\omega_K}\frac{e^{-i\omega u}}{\sqrt{4\pi\omega}}+\beta^{R}_{\omega,\omega_K}\frac{e^{i\omega u}}{\sqrt{4\pi\omega}}\right) \Theta(x)\right]
   \quad\,,
\end{equation}
which can be interpreted as positive-energy modes moving rightward in the R-region and their negative-energy partner-modes moving leftward in the L-region. The Bogolubov coefficients are evaluated to be:
\begin{subequations}
   \begin{align}
    \alpha^{L}_{\omega,\omega_K}&=\frac{(i\omega_K)^{-i\omega/\kappa}}{2\pi\kappa}\sqrt{\frac{\omega}{\omega_K}}\Gamma\left(\frac{i\omega}{\kappa}\right);\,\,\,\,\,\,\quad \beta^{L}_{\omega,\omega_K}=\frac{(i\omega_K)^{i\omega/\kappa}}{2\pi\kappa}\sqrt{\frac{\omega}{\omega_K}}\Gamma\left(\frac{-i\omega}{\kappa}\right)\\\alpha^{R}_{\omega,\omega_K}&=\frac{(-i\omega_K)^{i\omega/\kappa}}{2\pi\kappa}\sqrt{\frac{\omega}{\omega_K}}\Gamma\left(\frac{-i\omega}{\kappa}\right);\quad \beta^{R}_{\omega,\omega_K}=\frac{(-i\omega_K)^{-i\omega/\kappa}}{2\pi\kappa}\sqrt{\frac{\omega}{\omega_K}}\Gamma\left(\frac{i\omega}{\kappa}\right)
\end{align} 
\end{subequations}
Plugging this in, the correlators for the Unruh state are evaluated to be:

\begin{subequations}\label{eq:Unruhcorrs}
    \begin{align}
        \langle \{\hat{\varphi}(x,t),\hat{\varphi}(x',t')\}\rangle&=\int\frac{d\omega}{2\pi\omega}\left[f\left(\frac{\pi\omega}{\kappa}\right)\cos{[\omega(u-u')]}+\cos{[\omega(v-v')]}\right]\\
        \langle \{\hat{\varphi}(x,t),\hat{\pi}(x',t')\}\rangle&=\int\frac{d\omega}{2\pi}\left[f\left(\frac{\pi\omega}{\kappa}\right)\frac{\sin{[\omega(u-u')]}}{c(x')-v_0}+\frac{\sin{[\omega(v-v')]}}{c(x')+v_0}\right]\\
        \langle \{\hat{\pi}(x,t),\hat{\pi}(x',t')\}\rangle&=\int\frac{d\omega\,\omega}{2\pi}\left[f\left(\frac{\pi\omega}{\kappa}\right)\frac{\cos{[\omega(u-u')]}}{\left(c(x)-v_0\right)\left(c(x')-v_0\right)}+\frac{\cos{[\omega(v-v')]}}{\left(c(x)+v_0\right)\left(c(x')+v_0\right)}\right],\\
        \text{where}\quad f\left(\frac{\pi\omega}{\kappa}\right)&=\begin{cases}
        \coth\left(\frac{\pi\omega}{\kappa}\right) & \text{for L-L or R-R correlations} \\
        \csch\left(\frac{\pi\omega}{\kappa}\right) &  \text{for L-R correlations}
    \end{cases}
    \end{align}
\end{subequations}

\subsection{II. Regularization approach and entanglement negativity simulations}

\subsubsection{A. Standard approach: The harmonic lattice}

The Hamiltonian for a scalar field of mass $m_f$ in a 1+1D Minkowski background ($v_0\to 0$ limit of the PG metric) can be discretized into a harmonic chain by imposing a UV cutoff $\epsilon$ and an IR cutoff $L=(N+1)\epsilon$~\cite{1993SrednickiPRL}:
\begin{equation}
	\label{eq:1DHami-Mink}
	{H}=\frac{1}{2}\int d{x} \left[{\pi}^2+(\nabla{\varphi})^2+m_f^2\varphi^2\right]\to\frac{1}{\epsilon}\tilde{H}\quad ;\quad \tilde{H}=\frac{1}{2}\left[\sum_{j=1}^{N}\tilde{\pi}_j^2+\sum_{i,j=1}^{N}K_{ij}\tilde{\varphi}_i \tilde{\varphi}_j\right],
\end{equation}
where the rescaled Hamiltonian $\tilde{H}$ and the quadratures $\tilde{\pi},\tilde{\varphi}$ are all dimensionless. Depending on the boundary conditions, the coupling matrix $K_{ij}$ becomes a symmetric Toeplitz matrix with the following nonzero elements:
	\begin{equation}
			K_{j,j+1}=K_{j+1,j}=-1\,,\quad K_{jj\neq 1 ,N}=2+m_f^2\epsilon^2\,,\quad K_{jj= 1 ,N}=\begin{cases}
		    2+m_f^2\epsilon^2 & \text{Dirichlet BC}\\
            1+m_f^2\epsilon^2 & \text{Neumann BC}
		\end{cases}
	\end{equation}
The distribution of normal modes (eigenvalues of $\vec{K}$) follows a dispersion relation in terms of wavenumbers $k_j$:
	\begin{eqnarray}\label{eq:dirmodes}
		k_j=\frac{j\pi}{L};\quad \tilde{\omega}_j^2=m_f^2\epsilon^2+\begin{cases}
		    4\sin^2{\left[\frac{ k_j\epsilon}{2}\right]} & \text{Dirichlet BC},\,\,j=1,..N\\
            4\sin^2{\left[\frac{ k_j\epsilon}{2}\left(1+\frac{1}{N}\right)\right]} & \text{Neumann BC},\,\,j=0,..N-1
		\end{cases}
\end{eqnarray}
The UV mode therefore corresponds to $\max{\{j\}}$, which as $N\to\infty$ asymptotes to the \textit{Nyquist} choice $k_{\rm UV}\to\pi/\epsilon$. For the vacuum case, the covariance matrix takes the following form:
\begin{equation}
    \mathbfs{\Sigma}=\frac{1}{2}\begin{bmatrix}       \vec{K}^{-1/2}&O\\O&\vec{K}^{1/2}&    \end{bmatrix}
\end{equation}
It can be seen that the covariance matrix is well defined provided the coupling matrix is positive semi-definite. The negativity scaling plotted in \ref{fig:VacuumNeg} includes the harmonic lattice case where a subsystem of $N=10^3$ oscillators are considered for bipartition, embedded within a system of $2\times10^3$ oscillators. This is analogous to considering an accessible subregion of interest within a much larger system, which hence corresponds to a mixed state for which negativity serves as an operational measure of entanglement (as opposed to the von Neumann entropy). For this plot we also considered Neumann BC setting $m_f\epsilon=2\times 10^{-4}$, so that the IR mode $\tilde{\omega}_0$ coincides with the IR cutoff we consider in the latter sections.  The primary issue with extending this approach to the supersonic region is that the coupling matrix $\vec{K}$ is no longer positive semidefinite. Nevertheless, it can still be extended to the black-hole exterior~\cite{1997Mukohyama.etalPRD,1998Mukohyama.etalPRD}, and serves as a standard numerical tool that can be adapted to various semi-classical settings~\cite{2007Das.ShankaranarayananCaQG,2020Chandran.ShankaranarayananPRD,2024Chandran.etalPRD,2025Belfiglio.etalPRDa,2026Chandran.Rajeev}.

\subsubsection{B. A new approach: Bandlimited correlations on a coarse-grained lattice}

The harmonic lattice constructed above is easy to handle --- it corresponds to a system with $N$ oscillator degrees of freedom and can be exactly described by a pure state --- however it asymptotes to the quantum field only in the continuum limit of $\epsilon\to0$ and $N\to\infty$. Besides the difficulty in extending to black-hole interiors, the harmonic lattice also does not faithfully capture the coarse-graining inherent to experimental detection. For the simplest implementation of the latter, we consider two free parameters: the lattice spacing $\epsilon$ that fixes the coarse-graining, and the bandwidth $k_{\rm UV}$ that fixes the spatial resolution in correlation measurements. For the Minkowski vacuum we see:
\begin{equation}
   \langle \{\hat{\varphi}(x,T),\hat{\varphi}(x',T')\}\rangle=\int_{0}^{\omega_{1}}\frac{d\omega}{2\pi\omega}\left[\cos{[\omega(u-u')]}+\cos{[\omega(v-v')]}\right]= \int_{-k_{\rm UV}}^{k_{\rm UV}} \frac{cdk}{2\pi \omega}\cos\left[\omega(T-T')\right]e^{ik(x-x')}\, ,
\end{equation}
where the dispersion relation $\omega_{1}=ck_{\rm UV}$ is satisfied by the incoming/outgoing UV modes. However in the PG vacuum this relation is asymmetric, and from \eqref{eq:vacuumcorrs} we see that the covariance matrix decomposes into respective contributions from outgoing/incoming modes, i.e., $\vec{\Sigma}=\vec{I}+\vec{J}$. The coarse-graining on the other hand amounts to filling up a finite-dimensional covariance matrix with correlators discretely sampled along $x_j=-\frac{L}{2}+j\epsilon$ where $j=0,...,N-1$ and $L=(N-1)\epsilon$ (exactly $N$ points along $x\in[-L/2,L/2]$). 
The lattice point exactly at the sonic horizon ($x=0$) is averted by imposing $L=N$ (the closest points are $x=\pm \epsilon/2$ where $\epsilon\geq1$). The resulting covariance matrix describes a mixed state from both coarse-graining and confining to a field subregion. Unlike the harmonic lattice, the sampled lattice does not violate the continuum nature of the field, and the coarse-graining is implemented at the level of the covariance matrix. To ensure dimensionless entries, we rescale the correlators (or equivalently, $\pi\to\epsilon\pi$) as:
\begin{equation}
    (\Sigma_{\varphi\varphi})_{ij}=\frac{1}{2}\langle \{\hat{\varphi}_i,\hat {\varphi}_j\}\rangle,
        \quad(\Sigma_{\varphi\pi})_{ij}=\frac{\epsilon}{2}\langle \{\hat {\varphi}_i,\hat{\pi}_j\}\rangle
        ,\quad 
        (\Sigma_{\pi\pi})_{ij}=\frac{\epsilon^2}{2}\langle \{\hat{\pi}_i,\hat{\pi}_j\}\rangle\,.
\end{equation}

The sampled lattice corresponds to a Nyquist lattice when $\epsilon\to\epsilon_{\rm Nyq}\equiv\pi/k_{\rm UV}$ --- which captures the most \textit{optimal} coarse-graining relative to the bandlimit imposed on correlation measurements. That is, if $\epsilon>\epsilon_{\rm Nyq}$ we do not extract the maximum possible information allowed by the bandwidth, whereas if $\epsilon<\epsilon_{\rm Nyq}$ we enter length scales that are not resolved by the bandwidth, leading to spurious features. For the main simulations, we incorporate the Nyquist choice by fixing the spectral cutoffs that respect $k_{\rm UV}=\pi/\epsilon$ for the incoming/outgoing contributions of two-point correlators. Our approach differs from \cite{2015Pye.etalPRD,2023Lewis.etalPRD} in that we coarse-grain the bandlimited correlations (the primary observables in the experiment) from the continuum, as opposed to employing an equivalent, discrete representation that lives fully on the lattice --- interestingly this leads to different results for the undersampling ($\epsilon >\epsilon_{\rm Nyq}$) and oversampling ($\epsilon <\epsilon_{\rm Nyq}$) scenarios, as discussed later in Sec.\,\,II.D. Although our approach prevents us from employing pure state measures such as entanglement entropy, it is the simplest implementation of the operational limits of the experimental setup, and also provides a broader blueprint for extracting entanglement measures from curved spacetime QFT settings.

For the PG vacuum, the covariance matrix elements are thus evaluated as follows:
\begin{subequations}
\begin{align}  
(I_{\varphi\varphi})_{ij}&=\int_{\tilde{\omega}_{0u}}^{\tilde{\omega}_{1u}}\frac{d\tilde{\omega}}{4\pi\tilde{\omega}}\cos{\left(\frac{\tilde{\omega}|i-j|}{1-\tilde{v}_0}\right)}=\frac{1}{4\pi}\bigg[\ci{\left(\frac{\tilde{\omega}_{1u}|i-j|}{1-\tilde{v}_0}\right)}-\ci{\left(\frac{\tilde{\omega}_{0u}|i-j|}{1-\tilde{v}_0}\right)}\bigg] \\
(J_{\varphi\varphi})_{ij}&=\int_{\tilde{\omega}_{0v}}^{\tilde{\omega}_{1v}}\frac{d\tilde{\omega}}{4\pi\tilde{\omega}}\cos{\left(\frac{\tilde{\omega}|i-j|}{1+\tilde{v}_0}\right)}=\frac{1}{4\pi}\bigg[\ci{\left(\frac{\tilde{\omega}_{1v}|i-j|}{1+\tilde{v}_0}\right)}-\ci{\left(\frac{\tilde{\omega}_{0v}|i-j|}{1+\tilde{v}_0}\right)}\bigg]
\end{align}
\end{subequations}
Similarly, the other integrals in $\vec{\Sigma}$ can be evaluated exactly. The Nyquist choice here corresponds to $\tilde{\omega}_{1u}=(1-\tilde{v}_0)\pi$ (for $\vec{I}$) and $\tilde{\omega}_{1v}=(1+\tilde{v}_0)\pi$ (for $\vec{J}$), satisfying the respective dispersion relations for incoming/outgoing UV modes. For the Unruh state, the covariance matrix elements constitute integrals that must be evaluated numerically:
\begin{subequations}
\begin{align}  
(I_{\varphi\varphi})_{ij}&=\int_{\tilde{\omega}_{0u}}^{\tilde{\omega}_{1u}}\frac{d\tilde{\omega}}{4\pi\tilde{\omega}}\,f_\omega(\tilde{\kappa})\cos{\left[\tilde{\omega}(\tilde{u}_i-\tilde{u}_j)\right]};\quad(J_{\varphi\varphi})_{ij}=\int_{\tilde{\omega}_{0v}}^{\tilde{\omega}_{1v}}\frac{d\tilde{\omega}}{4\pi\tilde{\omega}}\,\cos{\left[\tilde{\omega}(\tilde{v}_i-\tilde{v}_j)\right]}\\
(I_{\varphi\pi})_{ij}&=\int_{\tilde{\omega}_{0u}}^{\tilde{\omega}_{1u}}\frac{d\tilde{\omega}}{4\pi}\,\frac{f_\omega(\tilde{\kappa})\sin{\left[\tilde{\omega}(\tilde{u}_i-\tilde{u}_j)\right]}}{\tilde{c}_j-\tilde{v}_0};\,\,\,\quad (J_{\varphi\pi})_{ij}=\int_{\tilde{\omega}_{0v}}^{\tilde{\omega}_{1v}}\frac{d\tilde{\omega}}{4\pi}\,\frac{\sin{\left[\tilde{\omega}(\tilde{v}_i-\tilde{v}_j)\right]}}{\tilde{c}_j+\tilde{v}_0}\\
(I_{\pi\pi})_{ij}&=\int_{\tilde{\omega}_{0u}}^{\tilde{\omega}_{1u}}\frac{d\tilde{\omega}}{4\pi}\,\frac{\tilde{\omega}f_\omega(\tilde{\kappa})\cos{\left[\tilde{\omega}(\tilde{u}_i-\tilde{u}_j)\right]}}{\left(\tilde{c}_i-\tilde{v}_0\right)\left(\tilde{c}_j-\tilde{v}_0\right)};\quad (J_{\pi\pi})_{ij}=\int_{\tilde{\omega}_{0v}}^{\tilde{\omega}_{1v}}\frac{d\tilde{\omega}}{4\pi}\frac{\tilde{\omega}\,\cos{\left[\tilde{\omega}(\tilde{v}_i-\tilde{v}_j)\right]}}{\left(\tilde{c}_i+\tilde{v}_0\right)\left(\tilde{c}_j+\tilde{v}_0\right)}\,\,,\\
\text{where }\tilde{v}_0=\frac{v_0}{c_R},\quad \tilde{\omega}&=\frac{\omega \epsilon}{c_R},\quad \tilde{\kappa}=\frac{\kappa \epsilon}{c_R},\quad \tilde{c}_j=\frac{c(x_j)}{c_R},\quad f_{\omega}(\tilde{\kappa})=\begin{cases}
        \coth\left(\frac{\pi\tilde{\omega}}{\tilde{\kappa}}\right) & \text{for L-L or R-R correlations} \\
        \csch\left(\frac{\pi\tilde{\omega}}{\tilde{\kappa}}\right) &  \text{for L-R correlations}
    \end{cases}
\end{align}
\end{subequations}

\begin{figure}[!htb]
	\begin{center}
		\subfloat[][$\Sigma_{\varphi\varphi}$; $\tilde{v}_0=3/4$]{%
			\includegraphics[width=0.29\textwidth]{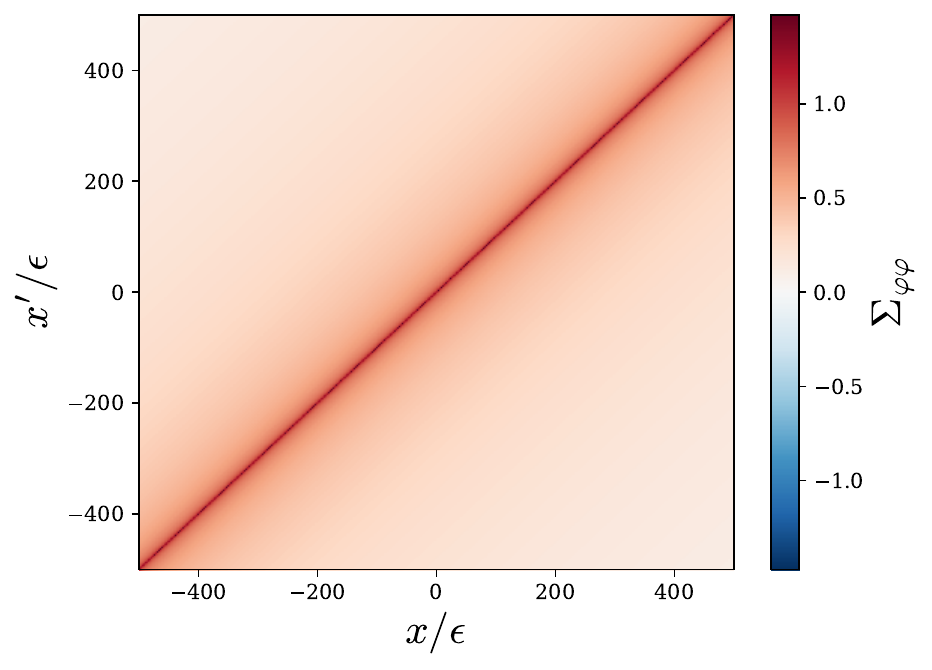}\label{fig:sxx0}
		}
		\subfloat[][$\Sigma_{\varphi\pi}$; $\tilde{v}_0=3/4$]{%
			\includegraphics[width=0.3\textwidth]{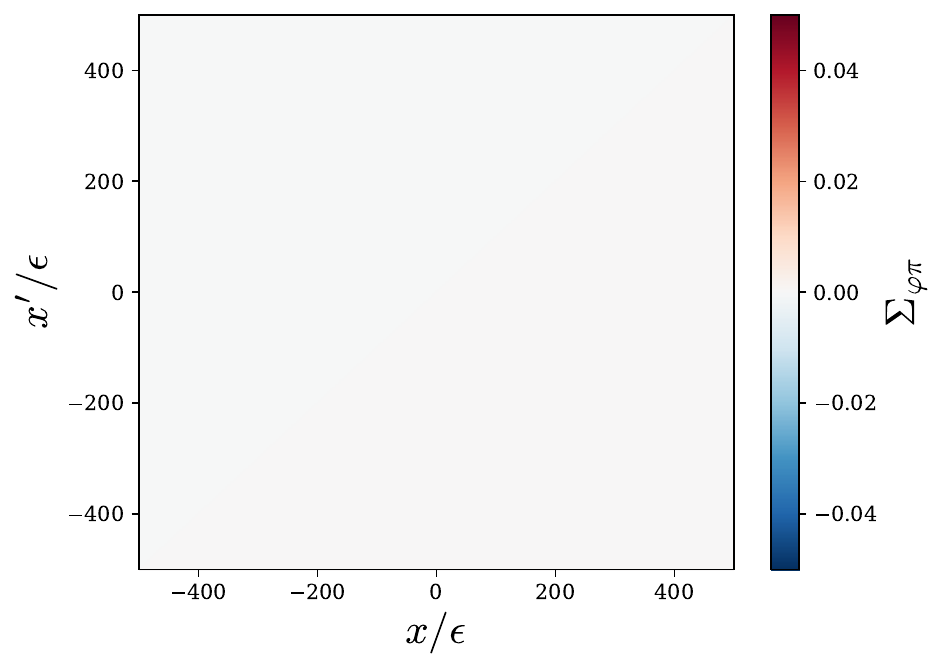}\label{fig:sxp0}
		}
        \subfloat[][$\Sigma_{\pi\pi}$; $\tilde{v}_0=3/4$]{%
			\includegraphics[width=0.3\textwidth]{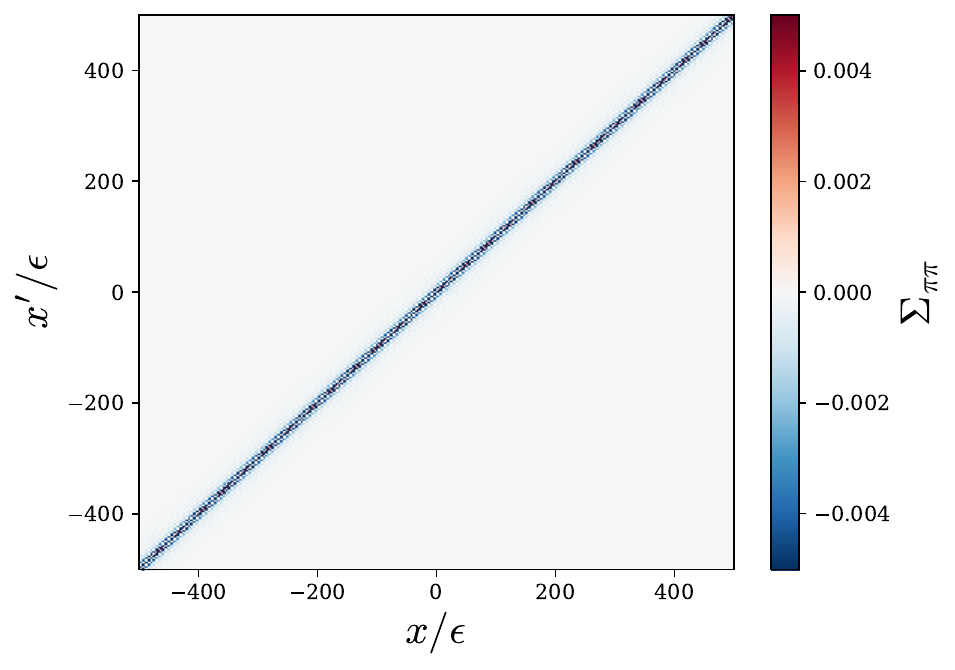}\label{fig:spp0}
		}
        \\
        \subfloat[][$\Sigma_{\varphi\varphi}$; $\tilde{v}_0=3/4$]{%
			\includegraphics[width=0.29\textwidth]{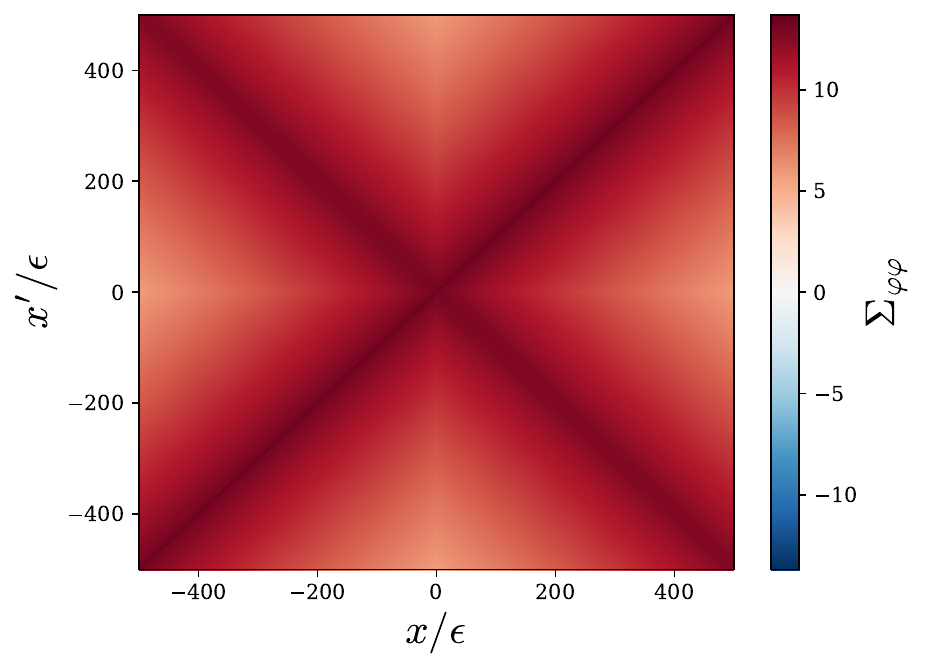}\label{fig:sxx1}
		}
		\subfloat[][$\Sigma_{\varphi\pi}$; $\tilde{v}_0=3/4$]{%
			\includegraphics[width=0.3\textwidth]{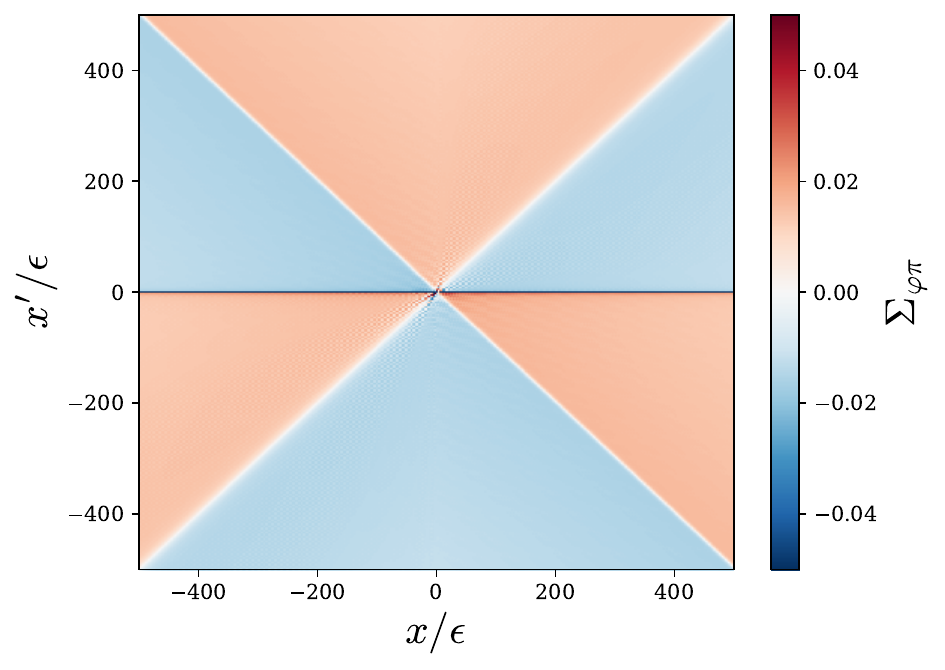}\label{fig:sxp1}
		}
        \subfloat[][$\Sigma_{\pi\pi}$; $\tilde{v}_0=3/4$]{%
			\includegraphics[width=0.3\textwidth]{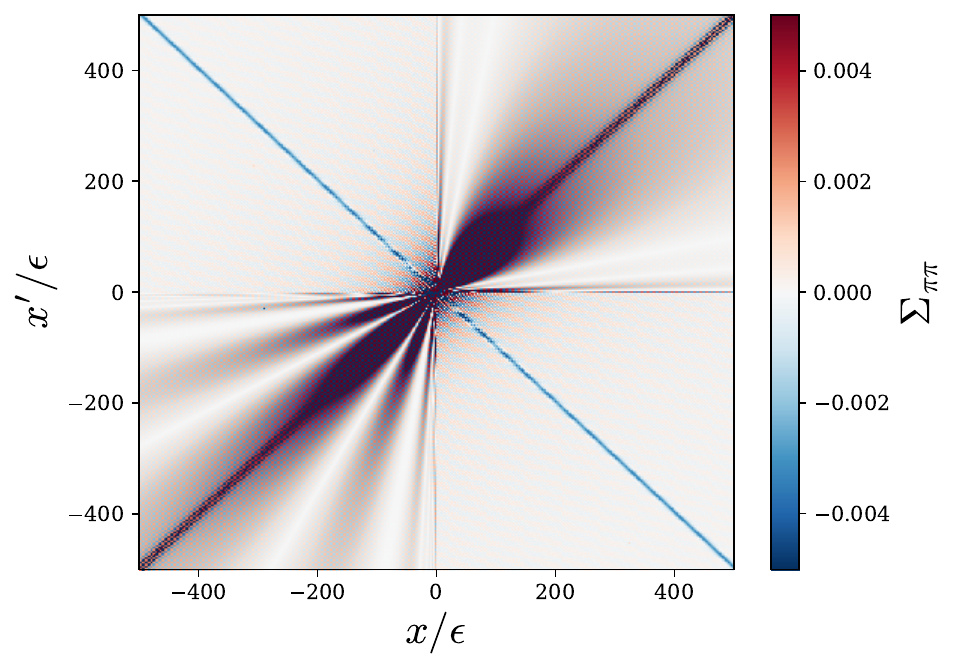}\label{fig:spp1}
		}
        \\
        \subfloat[][$\Sigma_{\varphi\varphi}$; $\tilde{v}_0=2/3$]{%
			\includegraphics[width=0.29\textwidth]{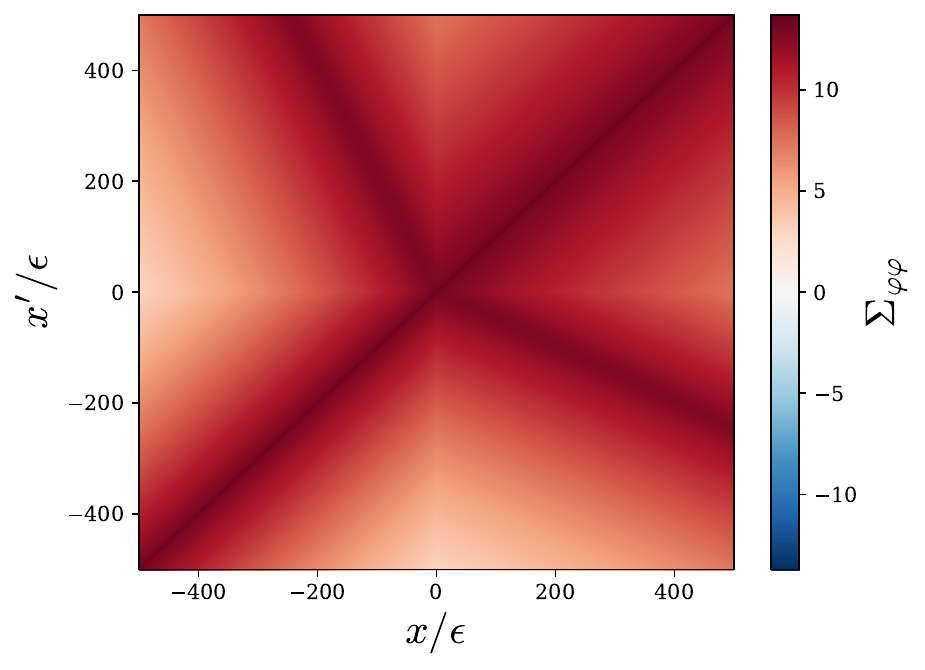}\label{fig:sxx2}
		}
		\subfloat[][$\Sigma_{\varphi\pi}$; $\tilde{v}_0=2/3$]{%
			\includegraphics[width=0.3\textwidth]{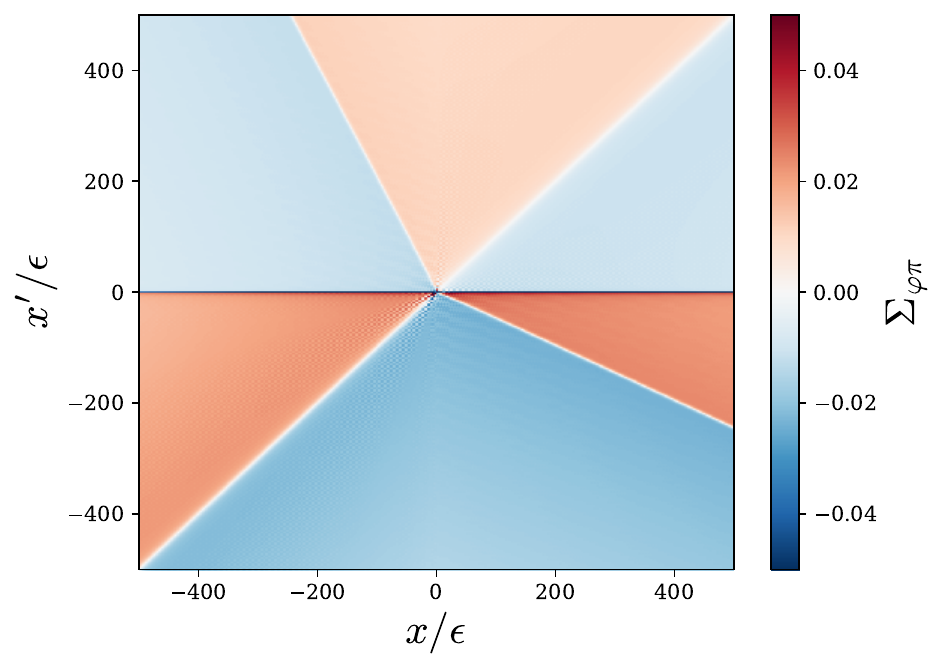}\label{fig:sxp2}
		}
        \subfloat[][$\Sigma_{\pi\pi}$; $\tilde{v}_0=2/3$]{%
			\includegraphics[width=0.3\textwidth]{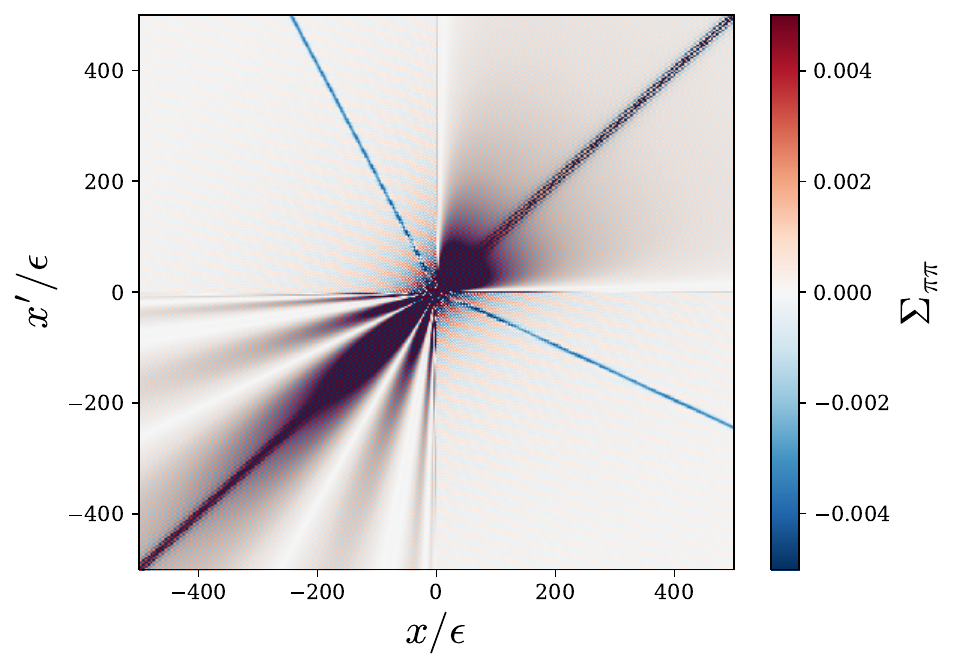}\label{fig:spp2}
		}
        \\
        \subfloat[][$\Sigma_{\varphi\varphi}$; $\tilde{v}_0=4/5$]{%
			\includegraphics[width=0.29\textwidth]{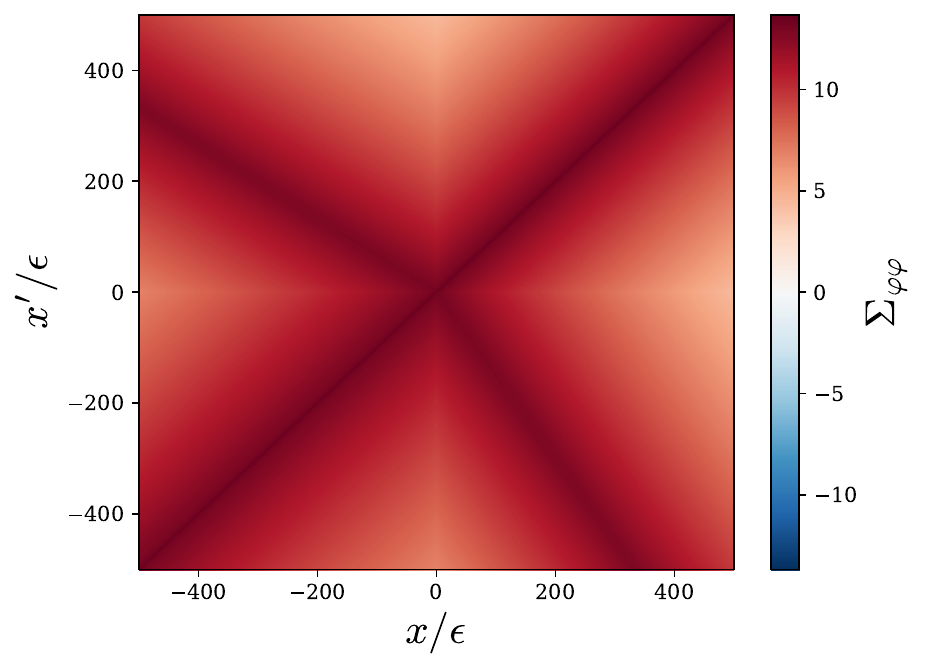}\label{fig:sxx3}
		}
		\subfloat[][$\Sigma_{\varphi\pi}$; $\tilde{v}_0=4/5$]{%
			\includegraphics[width=0.3\textwidth]{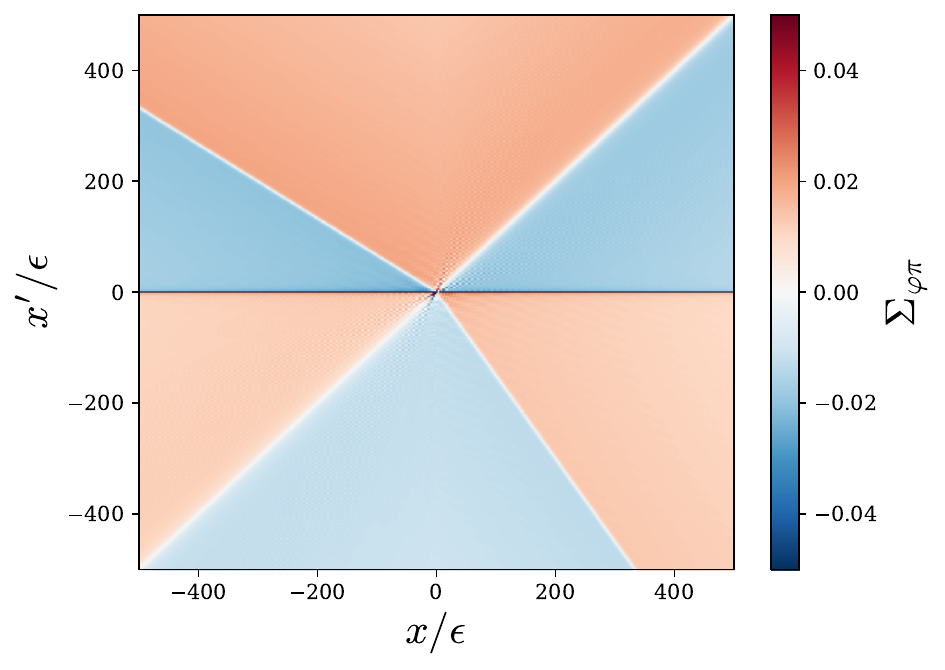}\label{fig:sxp3}
		}
        \subfloat[][$\Sigma_{\pi\pi}$; $\tilde{v}_0=4/5$]{%
			\includegraphics[width=0.3\textwidth]{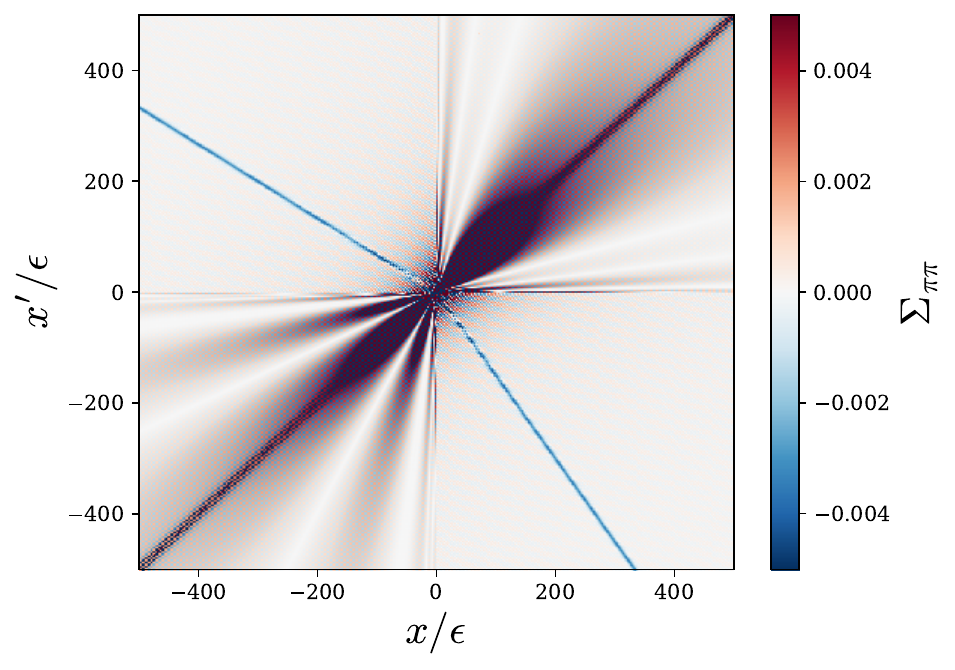}\label{fig:spp3}
		}
		\caption{Equal-time (T) correlators for the PG conformal vacuum in the absence of a horizon (a-c), and for the Unruh state in the presence of a horizon (d-l). For the latter, we have chosen $\tilde{\kappa}=0.1$ and $\tilde{c}_L=0.5$.}
		\label{fig:Sigma}
	\end{center}
\end{figure}

The (dimensionless) null coordinates are evaluated as $\tilde{u}_j=\frac{c_R}{\epsilon}[t(x_j)-x^*(x_j)]$ and $\tilde{v}_j=\frac{c_R}{\epsilon}[t(x_j)+x^*(x_j)]$, using the definition from \eqref{eq:coord1}. We fix the integration limits as per \cite{2013Anderson.etalPRD}, i.e., $X_1=-X_2=\epsilon$, whereas $X_3$ and $X_4$ are fixed such that $x^*_j(x=\epsilon)=\frac{\epsilon}{c_R}$ and $x_j^*(x=-\epsilon)=\frac{\epsilon}{c_R}$, and the IR cutoff is fixed as $\tilde{\omega}_{0u}=\tilde{\omega}_{0v}=2\times10^{-4}$. As for the UV cutoff, we implement the Nyquist choice that incorporates the incoming/outgoing mode velocities in the subsonic/supersonic regions. This corresponds to choosing $\tilde{\omega}_{1u}^L=(\tilde{v}_0-\tilde{c}_L)\pi$, $\tilde{\omega}_{1v}^L=(\tilde{v}_0+\tilde{c}_L)\pi$ for L-L correlations, and $\tilde{\omega}_{1u}^R=(1-\tilde{v}_0)\pi$, $\tilde{\omega}_{1v}^R=(1+\tilde{v}_0)\pi$ for R-R correlations. Although such a choice seems ambiguous for L-R correlations, the UV cutoff here is inconsequential due to the absence of a coincidence limit, and has no bearing especially outside the quantum atmosphere (beyond which the nonlocal Hawking correlations dominate). For simulation purposes we choose a harmonic mean of the corresponding UV modes, i.e., $\tilde{\omega}_{1u}=2/(1/\tilde{\omega}_{1u}^L+1/\tilde{\omega}_{1u}^R)$ and $\tilde{\omega}_{1v}=2/(1/\tilde{\omega}_{1v}^L+1/\tilde{\omega}_{1v}^R)$. We note that our results are insensitive to this choice, and any non trivial effects arising from it are negligible and confined to the quantum atmosphere. In \ref{fig:Sigma}, we henceforth present the simulated correlation structure for both the vacuum and Unruh states, showing the emergence of nonlocal peaks that signal Hawking pair production.

\subsubsection{C. Volume law fits for HR}

Using the scaling simulations from \ref{fig:ENsim} as a guide, we ascribe the following terms in the entanglement negativity for the Unruh state, up to an additional constant:
\begin{equation}\label{eq:fitting}
    \mathcal{E}_N\sim \frac{C_1}{4}\ln{\left[\frac{l_Al_B}{(l_A+l_B)\epsilon}\right]}+C_2\frac{l_H}{\epsilon}-C_3\frac{|l_A-l_H|}{\epsilon}\,,
\end{equation}
where the first term corresponds to the vacuum scaling, and the last two terms reflect the volume scaling observed about $x=0$. Here, $l_H$ corresponds to the subsystem size $l_A$ which coincides with the horizon at $x=0$, which in the employed bipartition scheme is fixed by $l_H=L/2=N\epsilon/2$. We treat the interior ($l_A<l_H$) and exterior ($l_A>l_H$) separately. 
To extract $C_1$ and $C_3$, the fitting function \eqref{eq:fitting} is applied while leaving out the near horizon (quantum atmosphere) region --- this is done by discarding 50 lattice points on either side of $l_A=l_H$. We also define $v_{\rm H}^L=v_0-c_L$ and $v_{\rm H}=c_R-v_0$ corresponding to the outgoing mode velocities in the interior and exterior regions respectively. When $v_{\rm H}^L\neq v_{\rm H}^R$, the volume law transitions to vacuum scaling for a subsystem range in the region with the lower $v_{\rm H}$ --- this is because for the (slower) Hawking particles propagating in this region, their (faster) corresponding partners lie beyond the bipartition boundary (i.e., beyond $x=L/2$), therefore not adding to the negativity content. In such cases, we extract $C_3$ via a direct linear fit with respect to $l_A$, \textit{only} for the range of values where the volume law is present (for instance, we consider $n_A\in[160,250]$ for the volume law fit in the L-region when $\tilde{v}_0=2/3$ and $n_A\in[350,495]$ in the R-region when $\tilde{v}_0=2/3$ w.r.t \ref{fig:ENsim}). To fix $C_2$, on the other hand, we study the negativity at $l_A=l_H$ and how it scales with $\tilde{\kappa}$. Although this partition lies in the quantum atmosphere, we see that negativity is only offset by a constant that does not depend on $\tilde{\kappa}$, i.e., it is most likely dependent on the regulators/system size.

\begin{figure}[t]
	\begin{center}
		\subfloat[][$\tilde{v}_0=2/3$]{%
			\includegraphics[width=0.31\textwidth]{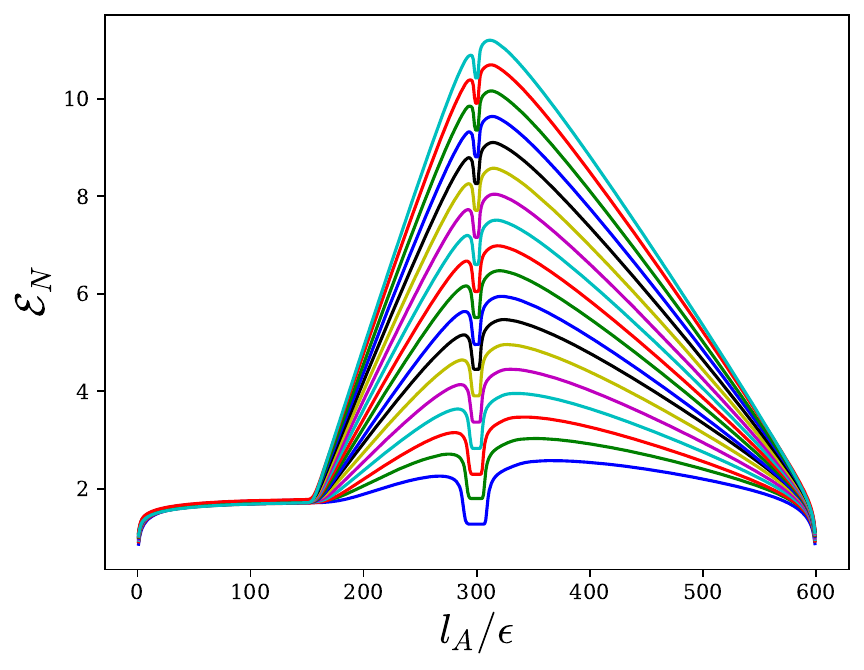}\label{fig:cv1}
		}
		\subfloat[][$\tilde{v}_0=3/4$]{%
			\includegraphics[width=0.31\textwidth]{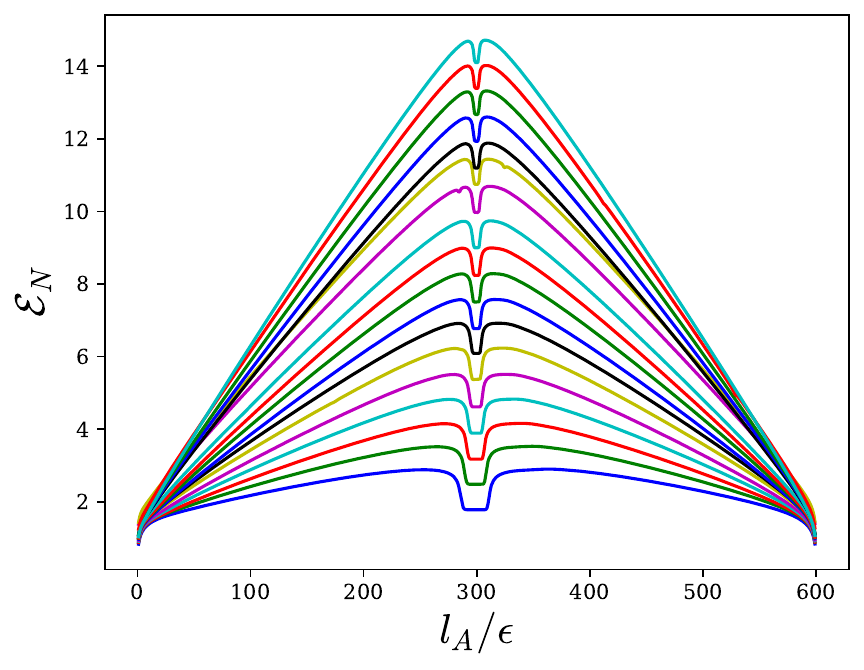}\label{fig:cv0}
		}
        \subfloat[][$\tilde{v}_0=4/5$]{%
			\includegraphics[width=0.31\textwidth]{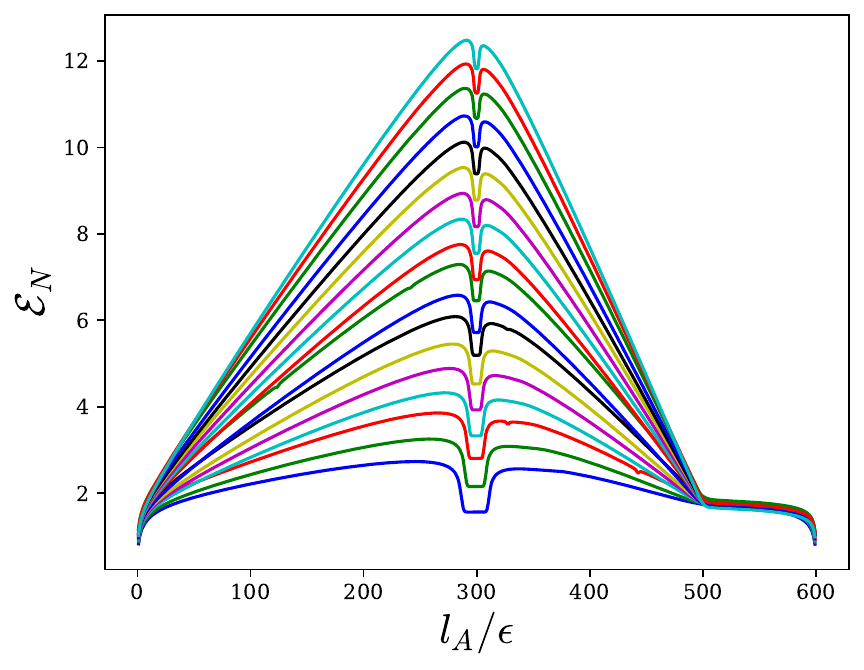}\label{fig:cv2}
		}

		\caption{Scaling of logarithmic negativity ($\mathcal{E}_N$) with subsystem size $n_A$ for increasing values of surface gravity in the range $0.01\leq\tilde{\kappa}\leq0.1$ (at intervals of $0.005$) and flow velocity $\tilde{v}_0$. Here, $N=600$ and $\tilde{c}_L=0.5$.}
		\label{fig:ENsim}
	\end{center}
\end{figure}

From \ref{fig:ENfits}, we see that both $C_2$ and $C_3$ are linear in $\tilde{\kappa}$, revealing that these terms are regulator-independent ($\frac{\tilde{\kappa}l_A}{\epsilon}=\frac{\kappa l_A}{c_R}$), and therefore, \textit{UV-finite}. $C_1$ on the other hand exhibits a nontrivial behavior --- it drops from the vacuum value of unity, asymmetrically in the interior and exterior regions. This points to potential UV-sensitive terms arising from HR (separate from the UV-finite terms), that affect the leading order vacuum scaling. Since these terms, collected as $\mathcal{E}_N^{\rm (UV)}$, are sensitive to the regulator $\epsilon$ and also become subleading for the coarse-grained lattice, we do not resolve them in this work. Instead we focus on the UV-finite terms arising from HR, collected as $\mathcal{E}_N^{\rm (HR)}$, as these are regulator-independent signatures that dominate the coarse-grained entanglement, as seen in \ref{fig:ENsim}.

\begin{figure}[t]
	\begin{center}
		\subfloat[][$C_1$]{%
			\includegraphics[width=0.305\textwidth]{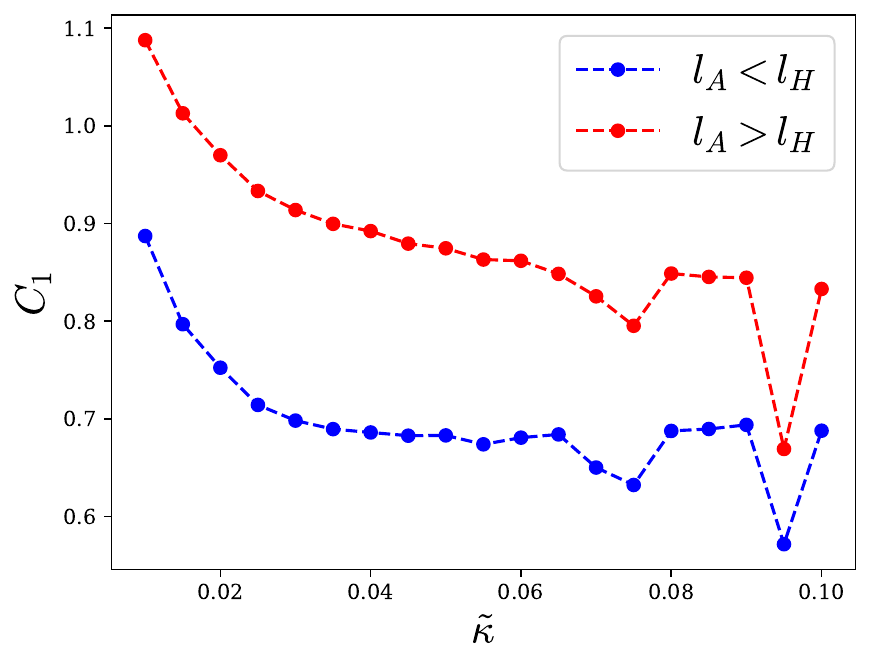}\label{fig:c1}
		}
		\subfloat[][$C_2$]{%
			\includegraphics[width=0.295\textwidth]{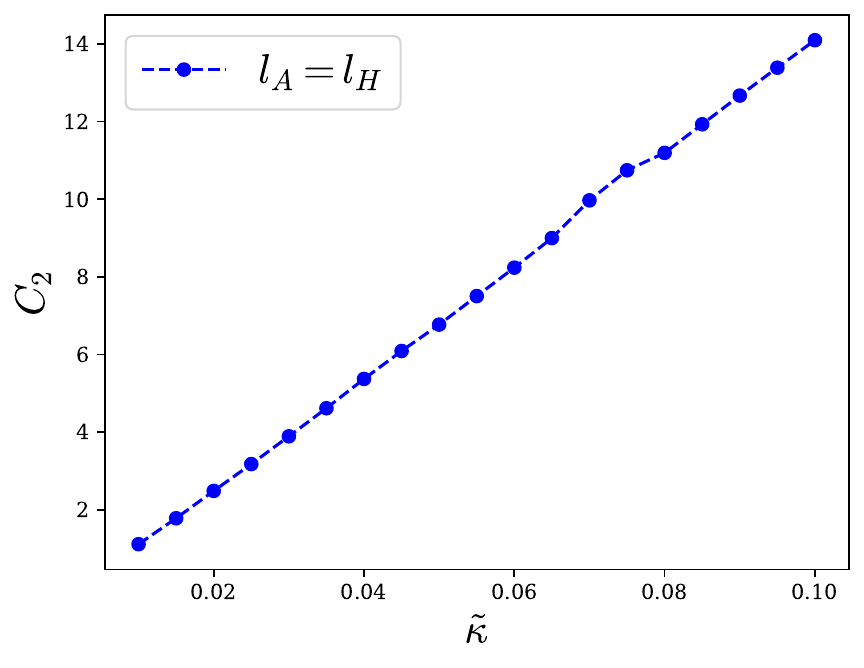}\label{fig:c2}
		}
        \subfloat[][$C_3$]{%
			\includegraphics[width=0.305\textwidth]{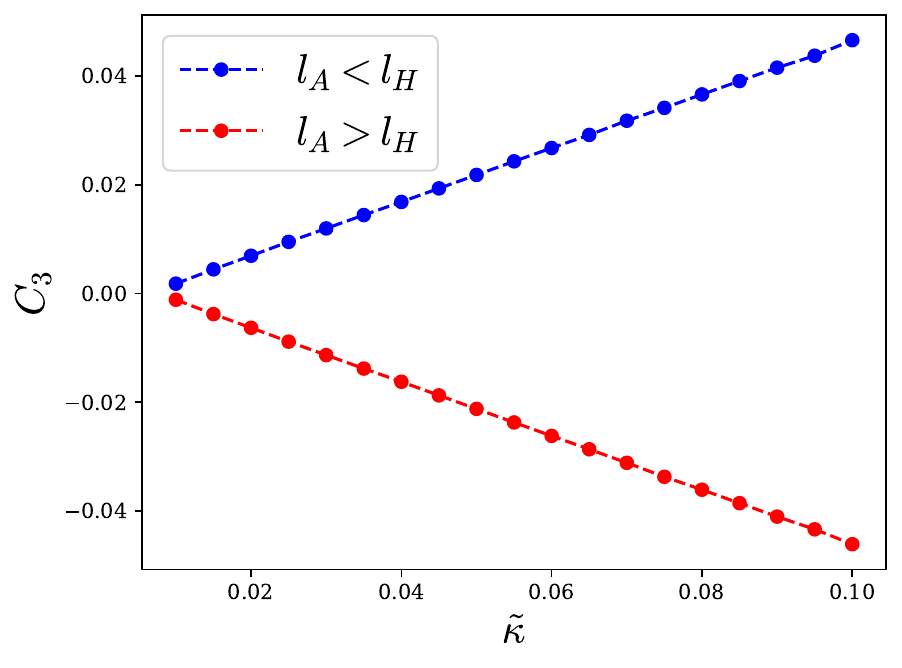}\label{fig:c3}
		}
        
		\caption{Prefactor fits for the predicted terms in logarithmic negativity and their dependence on surface gravity $\tilde{\kappa}$. Here, $N=600$, $\tilde{v}_0=3/4$ and $\tilde{c}_L=0.5$.}
		\label{fig:ENfits}
	\end{center}
\end{figure}
\begin{figure}[!htb]
	\begin{center}
        \subfloat[][$\tilde{v}_0=2/3$]{%
			\includegraphics[width=0.3\textwidth]{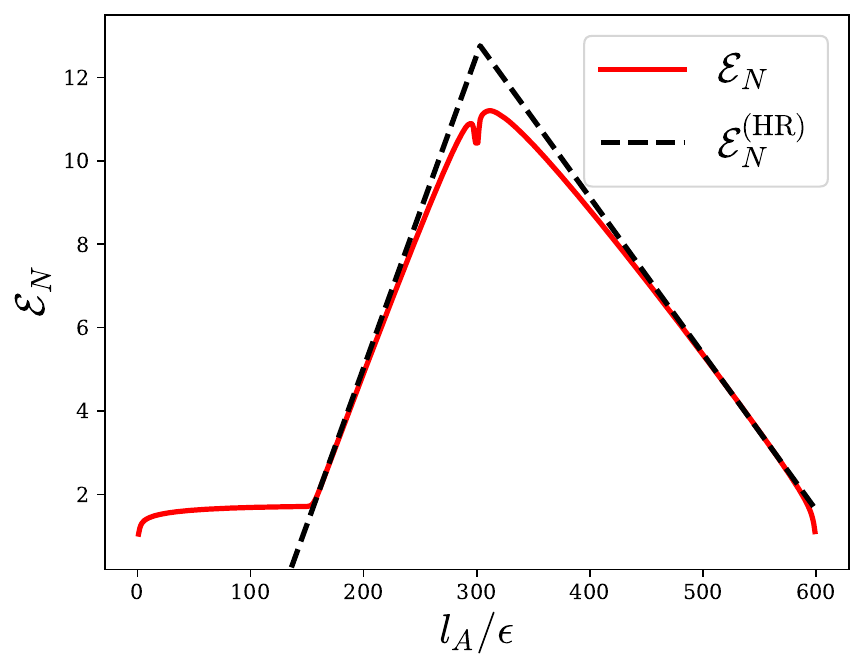}\label{fig:enfita}
		}
        \subfloat[][$\tilde{v}_0=3/4$]{%
			\includegraphics[width=0.3\textwidth]{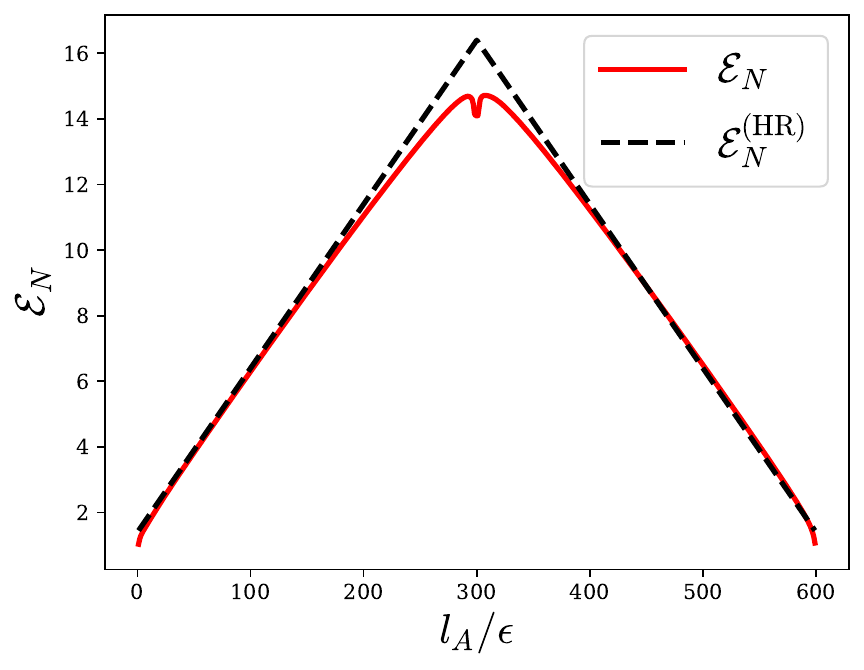}\label{fig:enfitb}
		}
        \subfloat[][$\tilde{v}_0=4/5$]{%
			\includegraphics[width=0.3\textwidth]{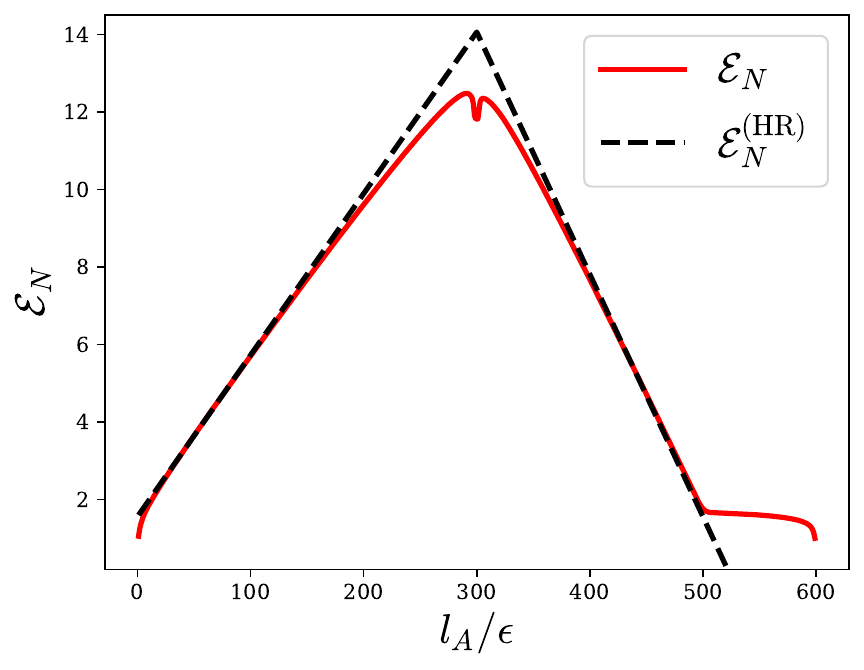}\label{fig:enfitc}
		}    
		\caption{Comparison of negativity scaling $\mathcal{E}_N$ with the UV-finite scaling of Hawking radiation $\mathcal{E}_N^{(\rm HR)}$. Here, $N=600$, $\tilde{\kappa}=0.1$ and $\tilde{c}_L=0.5$.}
		\label{fig:ENfits2}
	\end{center}
\end{figure}

The slopes of $C_2$ and $C_3$ from their linear fits with $\tilde{\kappa}$ are used to further fix these prefactors completely. Using the simulation data from \ref{fig:ENsim} and linear fits of \ref{fig:HR1}\subref{fig:HR1b}, we identify the prefactors to be $C_2={\kappa}/({8\max[v_{\rm H}^L,v_{\rm H}^R]})$, and $C_3={\tilde{\kappa}}/({8(1-\tilde{v}_0)})$ in the R-region while $C_3={\tilde{\kappa}}/({8(\tilde{v}_0-\tilde{c}_L)})$ in the L-region, within an error bar of $<2\%$. 
The scaling is therefore fixed, to a very good degree of accuracy outside the modulation region, as follows:
\begin{equation}
    \mathcal{E}_N\sim \mathcal{E}_N^{(\rm UV)}+\mathcal{E}_N^{(\rm HR)};\quad\quad  \mathcal{E}_N^{\rm (HR)}(l_A)\sim \frac{\kappa}{8}\left[\frac{l_H}{\max[v_{\rm H}^L,v_{\rm H}^R]}-\frac{|l_A-l_H|}{v_{\rm H}(l_A)}\right]\,,
\end{equation}
where $v_H(l_A<l_H)=v_{\rm H}^{L}$ and $v_H(l_A>l_H)=v_{\rm H}^{R}$. To see these fits more clearly, we superimpose them over the simulated scaling in \ref{fig:ENfits2} --- we see that fit is excellent away from the quantum atmosphere, but is overestimated within the quantum atmosphere. With increasing $\tilde{\kappa}$, the scaling further saturates the volume-law fit closer to the horizon as the quantum atmosphere shrinks as $\kappa^{-1}$~\cite{2021Fabbri.BalbinotPRL} --- however this can induce dispersive (analogue trans-Planckian) affects away from the hydrodynamical regime of the condensate. We can also in principle probe lower $\tilde{\kappa}$ values than the range considered --- but this requires the IR cutoff $\tilde{\omega}_{0}$ to be proportionately lower, and the bipartition region $L$ (and also $N$) to be significantly scaled up to go beyond the quantum atmosphere. Note that in the infinite boundary limit ($L\to\infty$), the scaling is divergent for near-horizon partitions but finite for far-away partitions, i.e., we get $\mathcal{E}_{N}^{\rm(HR)}\sim \kappa l_A/8v_{\rm H}^L$ in the far interior and $\mathcal{E}_{N}^{\rm(HR)}\sim \kappa l_B/8v_{\rm H}^R$ in the far exterior.

\subsubsection{D. Scaling robustness away from the Nyquist choice}
For our lattice-regularization framework, we considered two free parameters --- the lattice spacing $\epsilon$ that fixes the coarse-graining, and the bandwidth $k_{\rm UV}$ for correlation measurements in the laboratory. 
While the Nyquist coarse-graining corresponding to $\epsilon_{\rm Nyq}=\pi/k_{\rm UV}$ represents the optimal choice, we characterize the effects of oversampling ($\epsilon<\epsilon_{\rm Nyq}$) and undersampling ($\epsilon>\epsilon_{\rm Nyq}$) in \ref{fig:Noise}. We see that for $\epsilon<\epsilon_{\rm Nyq}$ we enter length scales not resolved by the correlation bandwidth, where the scaling picks up spurious and unregulated super-oscillations that one can interpret as ``UV noise". While the UV-finite term $\mathcal{E}_N^{\rm (HR)}$ remains intact, it becomes increasingly difficult to separate it from the noisy term $\mathcal{E}_N^{\rm (UV)}$ in the scaling. For $\epsilon >\epsilon_{\rm Nyq}$, the entanglement content is underestimated relative to what can be accessed by the bandwidth, fully dropping to zero for $\epsilon\geq 2\epsilon_{\rm Nyq}$ --- the coarse-graining should hence not exceed this bound. We also observe that the scaling behavior is largely preserved for the range $\epsilon_{\rm Nyq}\leq \epsilon \lesssim 2\epsilon_{\rm Nyq}$ --- this range is therefore favorable for extracting the results presented in this work. Note that these results differ from \cite{2015Pye.etalPRD,2023Lewis.etalPRD}, where a lattice representation for bandlimited quantum fields was studied. Most notably, a volume-law crossover was observed even for the vacuum state when $\epsilon>\epsilon_{\rm Nyq}$ (undersampled lattice), whereas in our approach the vacuum scaling is preserved up to $\sim 2\epsilon_{\rm Nyq}$ --- which may be chalked down to the fact that the correlators here do not live on the lattice, the covariance matrix is simply coarse-grained from the continuum. 

\begin{figure}[t]
	\begin{center}
        \subfloat[][$\epsilon<\epsilon_{\rm Nyq}$ (Vacuum)]{%
			\includegraphics[width=0.25\textwidth]{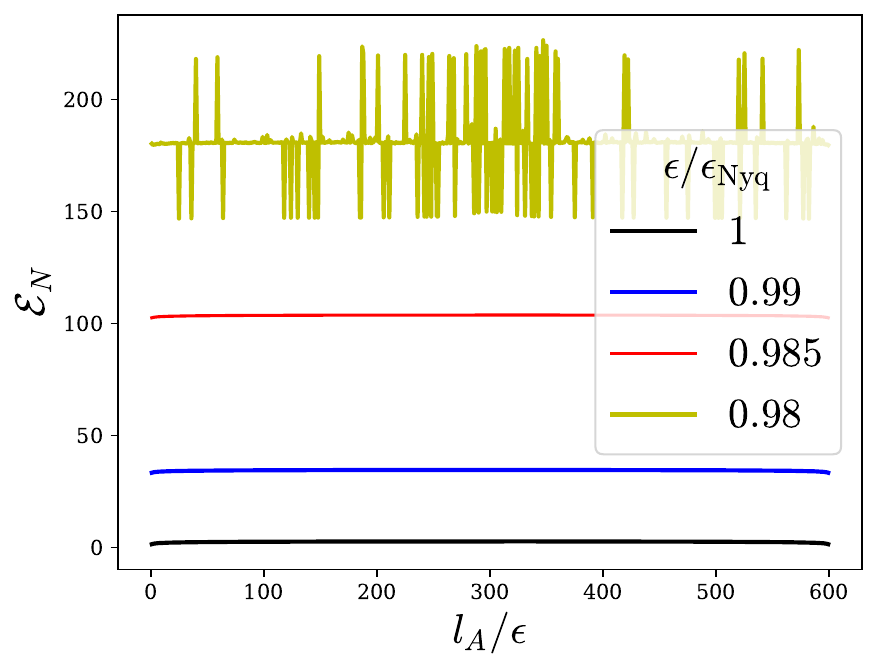}\label{fig:Nvaca}
		}
        \subfloat[][$\epsilon<\epsilon_{\rm Nyq}$ (HR)]{%
			\includegraphics[width=0.25\textwidth]{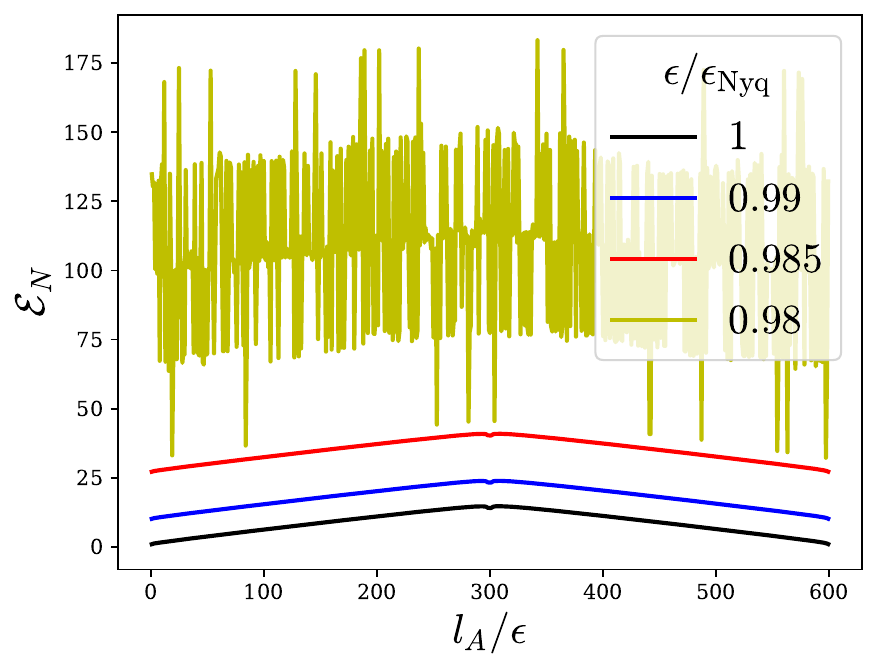}\label{fig:NHRa}
		}
        \subfloat[][$\epsilon>\epsilon_{\rm Nyq}$ (Vacuum)]{%
			\includegraphics[width=0.25\textwidth]{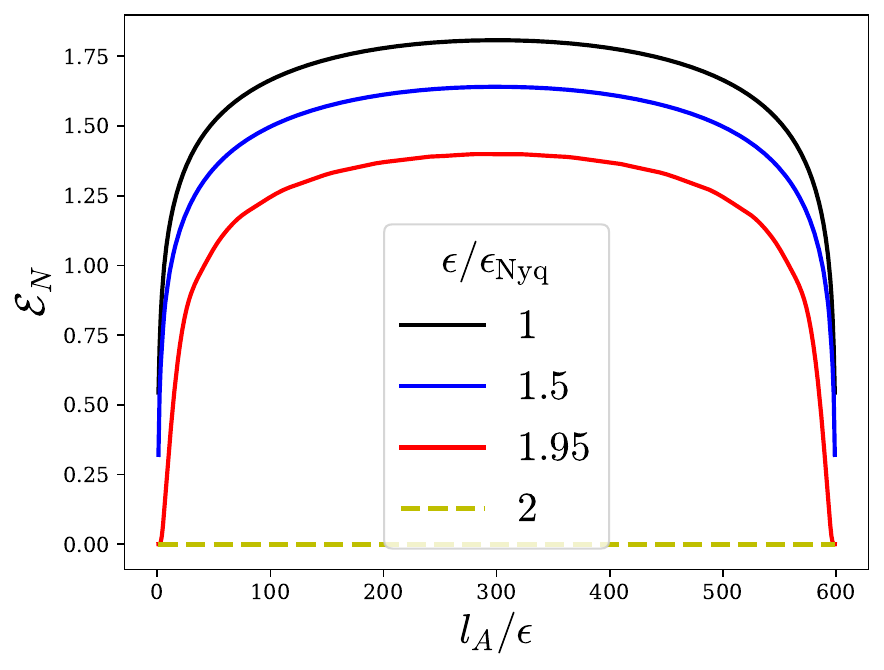}\label{fig:Nvacb}
		}      
        \subfloat[][$\epsilon>\epsilon_{\rm Nyq}$ (HR)]{%
			\includegraphics[width=0.25\textwidth]{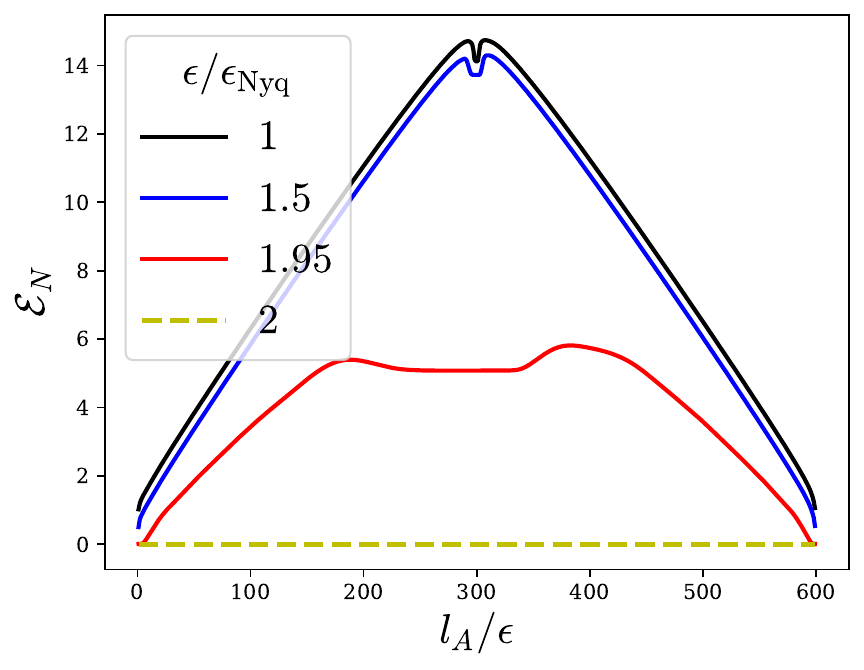}\label{fig:NHRb}
		}  
		\caption{Negativity scaling when the coarse-grained lattice is (a,b) oversampled or (c,d) undersampled relative to the bandwidth of correlation measurements. Here, $N=600$, $\tilde{v}_0=3/4$, $\tilde{\kappa}=0.1$, and $\tilde{c}_L=0.5$.}
		\label{fig:Noise}
	\end{center}
\end{figure}

\subsection{III. Comments on globally vs locally thermal scaling}
For a thermal CFT (which can also describe a thermal occupation state for the quasi-1D condensate), the negativity scaling for adjacent intervals ($l_A+l_B=L$) embedded in an infinitely large system is given below~\cite{2014Calabrese.etalJoPAMaT}:
\begin{equation}\label{eq:thermscaling}
    \mathcal{E}_N=\frac{1}{4}\ln{\left[\frac{\beta v_{\rm s}\sinh\left(\frac{\pi l_A}{\beta v_{\rm s}}\right)\sinh\left(\frac{\pi l_B}{\beta v_{\rm s}}\right)}{\pi \epsilon\sinh\left(\frac{\pi L}{\beta v_{\rm s}}\right)}\right]}+\rm const. \,,
\end{equation}
where $\beta=1/T$ is the inverse temperature, and $v_{\rm s}$ is the velocity of low-energy excitations. In the limit of small subsystem size $l_A/\beta v_{\rm s}\ll1$ and large embedding boundary $L/\beta v_{\rm s}\gg 1$, the scaling reduces to:
\begin{equation}\label{eq:thermlimit}
    \mathcal{E}_N\sim \frac{1}{4}\log\left(\frac{l_A}{\epsilon}\right)-\left(\frac{\pi T}{4 v_{\rm s}}\right) l_A+\mathcal{O}(l_A^2),
\end{equation}
where the UV-finite linear correction captures the degradation of entanglement with the global temperature $T$. Interestingly, the volume term we obtained for HR can be rewritten as:
\begin{equation}
    \mathcal{E}_{N}^{\rm (HR)}=\left(\frac{\pi T_{\rm H}}{4 v_{\rm H}^{\rm max}} \right)l_{H}-\left(\frac{\pi T_{\rm H}}{4 v_{\rm H}} \right)l_{\rm eff},
\end{equation}
where the Hawking temperature is given by $T_{\rm H}=\kappa/2\pi$, and $l_{\rm eff}=|l_A-l_H|$ is the interval length from the horizon. Interestingly, the second term is analogous to the volume correction arising in \eqref{eq:thermlimit}. At small distances from the horizon, this term (and only this term) can be interpreted as a thermal correction, i.e., the scaling appears \textit{locally thermal} for a partition near the horizon, somewhat in agreement with the conclusions of \cite{2025Sofos.etal}. However, the global effects are in fact the opposite, as thermal states cause a \textit{degradation} of vacuum entanglement~\cite{2014Calabrese.etalJoPAMaT}, whereas the Hawking effect, as we have shown, \textit{amplifies} the entanglement content relative to the vacuum case. 
For non-vacuum states in the laboratory, one would therefore encounter a competition between classical correlations that proliferate from an initial thermal occupation, and quantum correlations arising from entangled particles emitted by HR. Logarithmic negativity (unlike von Neumann entropy) can nevertheless separate these effects and address the robustness of entanglement content against thermal degradation --- the quantification of which we leave for future work. 

\end{widetext}
\end{document}